\documentstyle[preprint,eqsecnum,aps]{revtex}

\def	\be		{\begin{equation}}
\def	\ee		{\end{equation}}
\def	\beq		{\begin{eqnarray}}
\def	\eeq		{\end{eqnarray}}
\def	\simlt		{{\lower.5ex\hbox{$\;\buildrel{\mbox{\footnotesize$<$}}\over{\mbox{\footnotesize$\sim$}}\;$}}}
\def	\simgt		{{\lower.5ex\hbox{$\;\buildrel{\mbox{\footnotesize$>$}}\over{\mbox{\footnotesize$\sim$}}\;$}}}
\def	\rlarrows	{{\lower.6ex\hbox{$\;\buildrel{\mbox{\footnotesize$\longrightarrow$}}
				\over{\mbox{\footnotesize$\longleftarrow$}}\;$}}}
\def	\rarrow		{{\;\mbox{\footnotesize$\longrightarrow$}\;}}
\def    \bfgamma	{{\mbox{\boldmath$\gamma$}}}
\def  \lambdabar {{\mbox{$\lambda\!\!\!\!\!\stackrel{\mbox{$\bf\_$}}{\hphantom{0}\vphantom{_|}}$}}\!\!\!\phantom{\lambda}}
\def	\define	{\stackrel{\rm def}{=}}

\begin{document}

\title{\bf Effect of Coulomb collisions on time variations of the solar neutrino flux}

\author{Leonid Malyshkin$^1$ and Russell Kulsrud$^2$}

\address{Princeton University Observatory, Princeton NJ 08544, USA \\
	$^1$leonmal@astro.princeton.edu, $^2$rkulsrud@astro.princeton.edu}

\date{\today}


\maketitle

\begin{abstract}
We consider the possibility of time variations of the solar neutrino flux due to the radial motion of the 
Earth and neutrino interference effects. We calculate the time variations of the detected neutrino flux and 
the extent to which they are suppressed by Coulomb collisions of the neutrino emitting nuclei.

We treat the process of neutrino emission under the second quantization method. A neutrino emitting nucleus 
suffers random Coulomb accelerations in different directions. Thus, the positions and the velocities of the 
nucleus are different at different emission times. We properly calculate these emission times and emitted 
neutrino energies by making use of the method of stationary phase, which leads to the causality and 
Doppler shift equations. We average our results over the Holtsmark distribution of nuclear accelerations. 
To properly treat the collisions, it is necessary to simultaneously include in our analysis all other 
significant physical decoherence effects: the energy averaging and the averaging over the position of neutrino 
emission.

We find that the collisional decoherence is not important (in comparison with other decoherence effects) in 
the case of neutrino line, but it may be important in the case of the continuous neutrino spectrum (or for 
high energy resolution explorations of the internal structure of the line). We find that the collisional 
decoherence averages out the time variations for the neutrino masses 
$m_2^2-m_1^2\simgt 10^{-7}\,{\rm eV}^2\,{\cal E}^{3/2}_{\rm MeV}$.

A simple and clear physical picture of the time dependent solar neutrino problem is presented and qualitative 
coherence criteria are discussed. Exact results for the detected neutrino flux and its time variations are 
obtained for both the case of a solar neutrino line, and the case of the continuous neutrino spectrum with a 
Gaussian shape of the energy response function of the neutrino detector. We give accurate constraints on the 
vacuum mixing angle and the neutrino masses required for flux time variations to not be suppressed.

\vspace{0.3cm}\noindent Pac(s): 26.65.+t, 14.60.Pq, 96.60.Jw
\end{abstract}


\narrowtext


\section{Introduction}\label{INTRODUCTION}

To explain the discrepancy between the observed flux of solar neutrinos and the predicted emission
rate it has been postulated that the neutrinos have two or three different mass states, and that these mass 
states are different from the flavor states~\cite{BCDR_85,Be_86,BDW_88,B_89,BB_90,B_96,BKS_98}. 
This extension of the standard electroweak theory leads to oscillations between the electron neutrino
and the different neutrino species. A key role in the solution of the solar neutrino problem is played by 
the MSW theory~\cite{MS_85,W_78}. In this theory there is a spherical resonance region at the radius 
$r=r_{\rm res}$. We denote the region inside the resonance radius, $r<r_{\rm res}$, as {\it region~I} and the 
region outside the resonance radius, $r>r_{\rm res}$, as {\it region~II} (see Fig~\ref{FIGURE_PATHS}). 
Hereafter, we suppose the solar core to be inside region~I (i.~e.~inside the resonance radius 
$r=r_{\rm res})$. In this case all neutrinos are emitted in region~I and they cross the resonance region 
once to enter region~II on their way to the Earth (as shown in Fig~\ref{FIGURE_PATHS}). The other situation,
when the resonance region in inside the solar core, so that neutrinos emitted in region~II cross the resonance 
twice or not at all, presents complications in notations rather than in physics, and it is discussed in Appendix~A.

In our paper, we consider the case of two types of neutrino flavors and correspondingly
of two types of neutrino mass states. In this case for each neutrino energy there are two orthogonal WKB 
solutions of Dirac equation out of the resonance region, which we denote as $+$ {\it and} $-$ {\it neutrino 
states}. These neutrino states exist in regions~I and ~II, where the WKB approximation is valid, and they 
coincide with neutrino pure mass states in the vacuum, but they are not given for the resonance region, 
where the WKB approximation breaks down. 

In the adiabatic approximation the $+$ neutrino states are connected to each other as well as the $-$ states 
are connected. So, we say that there are two {\it neutrino paths}: $(+,+)$ and $(-,-)$ [for example, the $(+,+)$ 
path means that the neutrino is emitted in region~I as the $+$ state and is converted, while crossing the 
resonance, into the $+$ state in region~II]. However, if the adiabatic approximation breaks down, then the 
$+$ and $-$ states jump one into another at the resonance radius. As a result, in this case there are four 
neutrino paths: $(+,+)$, $(+,-)$, $(-,-)$ and $(-,+)$ (see Fig~\ref{FIGURE_PATHS}). 
Due to electron scattering and nonzero neutrino masses these neutrino paths gain different quantum 
mechanical phases during the propagation to the Earth and hence interfere with each other. 

There are {\it two neutrino eigenfunctions for each neutrino energy}, which are solutions of Dirac equation 
in the entire region of space (including the resonance region, where the WKB approximation is not valid). It is 
convenient to choose them so that they coincide with the $+$ and $-$ states in region~I (where 
neutrinos are emitted). In region~II (after the resonance crossing) the eigenfunctions are sums of the 
$+$ and $-$ states, whose coefficients are determined by resonance jump 
formulas~(\ref{NONADIABATIC_EIGENFUNCTIONS})--(\ref{JUMP_AMPLITUDE}) [see Fig~\ref{FIGURE_PATHS}]. 
For this choice the entire eigenfunctions are orthogonal. Each neutrino propagating to the Earth with a given 
energy can be represented as a linear combination of two eigenfunctions or as a linear combination of four 
neutrino paths (with only two independent coefficients and all others related to the jump formulas). For 
example, in the adiabatic approximation there is no jump, so, there are only two paths, the $(+,+)$ and $(-,-)$, 
and the two eigenfunctions correspond to these two paths (out of the resonance region).

The MSW effect results in a reduction of the average number of detected electron 
neutrinos~\cite{P_86} and also produces time variations of the neutrino flux. The time variations of 
the detected neutrino flux are due to interference between different neutrino paths and are sensitive to 
the neutrino masses and the mixing angle.
There are several physical effects that can destroy the coherence between neutrino paths and limit
the observability of flux time variations. The most significant effect is an averaging of the observed
neutrino flux over the effective detected energy band $\sigma$. This averaging wipes out a cross term between
two neutrino paths if $\sigma\simgt\sigma_t\define\hbar/\tau$, where $\tau$ is the difference of propagation 
times of these paths.
For a neutrino line, $\sigma$ is simply the energy width of the line, $\sigma\approx T$ 
(thermal broadening effect). In the case of a detection of the continuous neutrino energy spectrum 
the energy response function of our detection device must be sufficiently narrow if one wants to observe flux 
time variations. 

A second effect, which can destroy coherence, and which is equally important,
is the averaging of the detected neutrino flux over the region of neutrino emission inside the sun. 
This effect depends on the electron density in a region of the solar core and it is important even for the case of 
very small neutrino masses (see Sec.~\ref{AVERAGING_OVER_POSITION}). As we will see below, the averaging over 
the position of neutrino emission greatly simplifies all calculations because only two cross terms between 
four neutrino paths survive. (Actually one of these cross terms is much larger than another, so there is 
only one potentially important cross term! This cross term varies with the motion of the Earth and therefore it
gives time variations of the detected neutrino flux.)

A third effect that can potentially average out time variations of the neutrino flux is collisional decoherence. 
Electric forces acting on the emitting nucleus from its neighbors lead to a negligible Stark effect on a nuclear
scale but they also lead to the acceleration of the nucleus.
The collisional decoherence was first discussed by Nussinov in $1976$~\cite{N_76}, who used a coherence criterion 
$\tau\simlt\hbar/\Delta{\cal E}$ with $\Delta{\cal E}$ approximately equal to the electric potential due to 
neighboring ions. For the $0.862\,{\rm MeV}$ $^7$Be neutrino line Nussinov got 
$\tau\simlt 3\times 10^{-17} {\rm sec}$, although the theoretical basis is uncertain. 
In $1985$ Krauss and Wilczek~\cite{KWi_85} suggested the plasma full collision time as an 
approximate value of the coherence time. They got $\tau\simlt 10^{-15} {\rm sec}$. As we will see, this is 
a considerable overestimate for the coherence time. More recently, Loeb~\cite{L_89} considered a collisional Doppler 
spectral broadening $\langle|\Delta{\cal E}|\rangle={\cal E}\langle|\Delta V|\rangle/2c$. Loeb obtained a criterion 
$\tau\simlt {(2\lambdabar/a)}^{1/2}$, where $a={\langle 1/V\nu_{col}\rangle}^{-1}$ was the harmonically averaged 
nuclear acceleration and $\nu_{col}$ was the collision frequency.
Loeb's coherence time for the $0.862\,{\rm MeV}$ $^7$Be neutrino line was $5\times 10^{-17} {\rm sec}$. 

Because of the uncertainty in these estimates of the criterion for the collisional decoherence, we felt a more
detailed analysis was desirable. In this paper we carry out such an analysis and derive the correct collisional 
decoherence criterion. It turns out that Nussinov's and Loeb's estimates are close to this criterion 
if the acceleration they used is replaced by the Holtsmark acceleration.

In most of the previous papers the behavior of the neutrinos is treated as a one-dimensional problem, where a 
constant neutrino energy is assumed and the evolution of the neutrino in space (or equivalently in time) is 
followed as it propagates to the Earth. The neutrino is assumed to be emitted into a single state in the high
density solar core. It propagates to the resonant region where it nonadiabatically splits into two states. 
These two propagate with different momenta to the Earth, and at the detector on the Earth these two states can 
interfere according to the accumulated phase difference during this latter propagation.

This procedure is adequate for treating the decoherence due to the energy averaging. However, it is not adequate
to properly treat the collisional decoherence of the neutrino flux time variations, since the neutrino emitting
nucleus moves erratically in time (suffering random accelerations in different directions). Thus, the positions
and the velocities of the nucleus are different at different neutrino emission times.  These emission times have to
be properly defined and evaluated. To do this, both space and time must be introduced into the problem.

In our calculations we follow the prescription of Fermi. We assume two definite spatial eigenfunctions for each 
neutrino energy (because we assume two neutrino flavors). We then expand the neutrino wavefunction in terms of 
these eigenfunctions. The components of this expansion are functions of time alone, as the neutrino is emitted by 
the moving nucleus. The components themselves have probability distributions, and after second quantization
these distributions are, in turn, expanded into two oscillator states corresponding to occupation or non-occupation
of the neutrino eigenfunctions. Finally, the differential equations in $t$ and $x$ for the time evolution 
(during emission) of the coefficients $a$ of the latter expansion (i.~e.~of the oscillator occupation amplitudes)
are derived and integrated.

For a neutrino line the decaying nucleus emits a neutrino with the fixed energy in the frame co-moving 
with the nucleus. By integrating the differential equations one finds that each of oscillator occupation 
amplitudes $a$ is an integral over a time parameter $t'$ corresponding to ``the time of emission''. 
Then we sum over excited oscillator states by integration over a spread of neutrino energies $\cal E$ in 
the laboratory frame and summing over $\pm$ indices (the summing over $\pm$ is actually the summing over 
two eigenfunctions for each energy, which coincide with the $+$ and $-$ adiabatic states at the position 
of emission). Thus, the emitted neutrino wavefunction is a double integral over $t'$ and $\cal E$ and, of 
course, a function of $t$ and $x$. Because of the rapidly changing phase the integrands in this double 
integral vary rapidly with $t'$ and $\cal E$. For a given time and position of detection $t$ and $X_{\rm E}$, 
the main contribution to the integral comes from isolated values for $t'$ and $\cal E$ where the phase is 
stationary. As one would expect, these values coincide with the solutions of the causality and Doppler shift 
equations for the emission time $t'$ and the emitted neutrino energy $\cal E$. The method of stationary phase 
selects these solutions and also gives the neutrino phase.

Let us, for simplicity, consider that the neutrino is emitted into only one of two neutrino eigenfunctions, 
for example, into the one that coincides with the $-$ state in region~I (as in case of high 
electron density in the solar core). Then, in region~II this eigenfunction is given by the sum 
of $-$ and $+$ states due to nonadiabatic resonance crossing. In other words, the entire eigenfunction 
is the mixture of two neutrino paths $(-,-)$ and $(-,+)$ out of the spherical resonance region 
(i.~e.~in regions~I and ~II). These two neutrino paths have the same momenta $k_-$ in region~I but 
different momenta $k_-$ and $k_+$ in region~II [$k_-<k_+$, see Eqs~(\ref{K_PRIME_DEFINITION}) 
and~(\ref{K_PRIME})]. Therefore, for a given time and position of the detection, one finds two solutions for 
$t'$ and $\cal E$ from the method of stationary phase, which correspond to the two different propagation 
times of the two neutrino paths. We can say that the eigenfunction is emitted twice: first the $(-,-)$ path 
contributes, then the $(-,+)$ path contributes.
The two different values of the emission time $t'$ are determined as the two retarded times. The corresponding two 
different values of the neutrino energy $\cal E$ are found to be the two Doppler-shifted energies, since the 
emitting nucleus has two different velocities at the two emission times due to its acceleration. In terms of these
parameters, $t'$ and $\cal E$, the relative phase of the two neutrino paths (i.~e.~of the two solutions for
the emission of the eigenfunction) can be found and the amount of interference between them can be determined.
Then we have to average over the distribution of nuclear accelerations, since we detect many emitting nuclei
and integrate over a large interval of detection time. Only in this way can we correctly determine the decoherence 
produced by collisions. The resulting decoherence depends on neutrino masses and the mixing angle.

For the continuous neutrino spectrum one needs to appreciate that an electron is emitted as well as a neutrino.
To carry out the problem in this case one, first, fixes the state into which the electron is emitted and then 
treats the neutrino emission as in the case of neutrino line emission. However, at the end a sum over 
electron states must be taken. This introduces some further complications. For example, the two neutrino paths
have different energies, so if the energy band width of the detector is too narrow, the detector will be affected
only by that one of these neutrino paths which has its energy inside the detector band, and the other path will be 
outside this energy band. Thus, no interference can occur. In the case of the continuous neutrino spectrum detection 
both collisional decoherence and averaging over the neutrino energy must be treated simultaneously.

In this paper we present a complete self-consistent analysis (making use of the method of stationary phase) 
of the time variations of the detected neutrino flux due to interference effects.
We consider a general nonadiabatic theory for two neutrino flavors. We neglect spherical geometry of the 
problem and use jump formulas derived under the assumption that the solar electron density varies exponentially 
with the radius throughout the resonance region. We make use of the last assumption only for our final numerical 
results. Another assumption that we make is that of constant nuclear acceleration, i.e.~the acceleration of the 
emitting nucleus is considered constant during the delay time $\tau$ between propagations of different neutrino 
paths. However, the nuclear acceleration has random direction. We use a Holtsmark distribution for the absolute 
value of the instantaneous nuclear acceleration~\cite{Ch_43}. The assumption of constant nuclear acceleration 
turns out to be sufficient for our purposes, i.e.~it is valid for all relevant cases in which time variations of 
the detected neutrino flux are significant.

\begin{figure}
\vspace{9.2cm}
\includegraphics{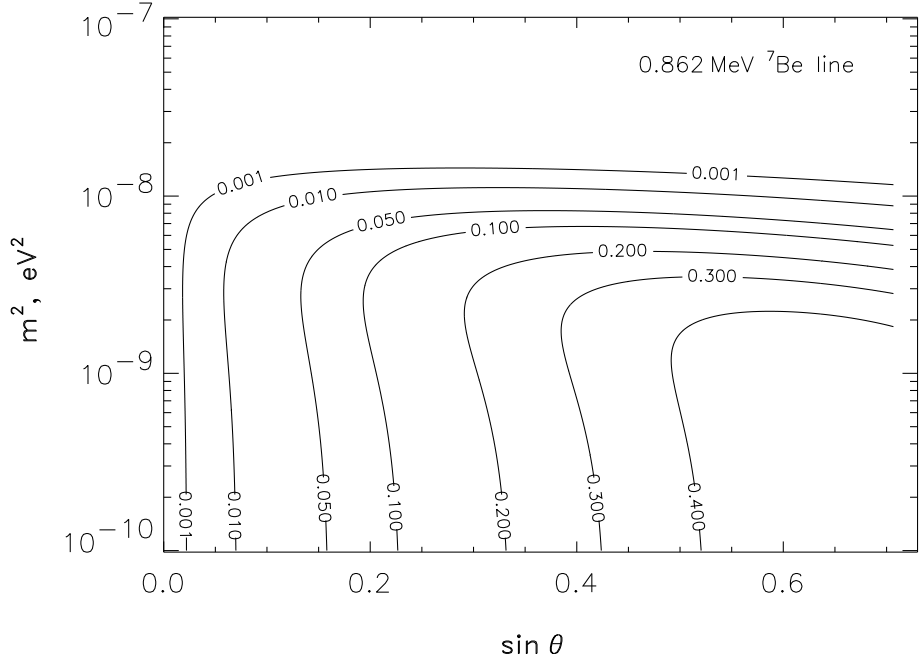}
\caption{
Contours $A_{\rm MSW}={\rm const}$ are shown for the $0.862\,{\rm MeV}$ $^7$Be neutrino line. 
These curves represent the amplitude of neutrino flux time variations normalized to the standard neutrino flux.
If we neglect nuclear acceleration, the curves will not noticeably change.}
\label{FIGURE_LINE_ENERGY}
\end{figure}

\begin{figure}
\vspace{9.0cm}
\includegraphics{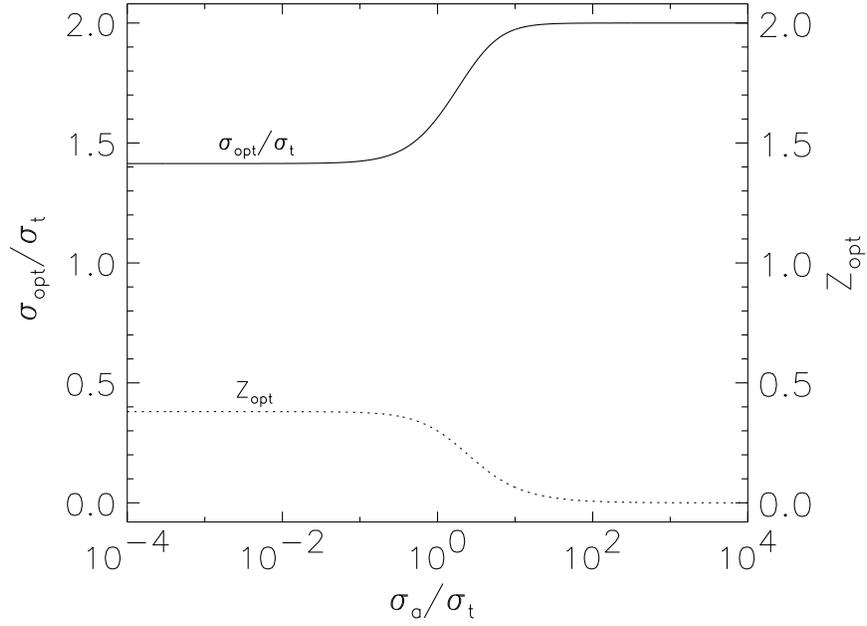}
\caption{
The optimal value of parameter $\xi=\sigma/\sigma_t$ as a function of the parameter $\alpha=\sigma_t/\sigma_a$ is shown 
by the solid line. The corresponding maximum value of the function $Z(\xi,\alpha)$ is shown by the dotted line.}
\label{FIGURE_CONTINUOUS_OPTIMAL}
\end{figure}

In the second section we use the WKB approximation to solve the Dirac equation in the mass state representation
and to derive formulas for the $+$ and $-$ neutrino states. Then, in Sec.~\ref{JUMP} we discuss the nonadiabatic 
jump at the resonance radius and we write down the two neutrino eigenfunctions. In Sec.~\ref{SECOND_QUANTIZATION} 
we study the processes of neutrino emission and detection under the second quantization theory. Using the method 
of stationary phase, we obtain the neutrino wave function at the Earth and the causality and Doppler shift 
equations in Sec.~\ref{STATIONARY_PHASE}. These equations are solved and the quantum mechanical phase differences 
between neutrino paths are calculated in Sec.~\ref{CAUSALTY_DOPPLER} making use of an expansion in small 
neutrino masses. Since the final exact calculations are quite complicated, we give a qualitative analysis of 
the problem in Sec.~\ref{QUALITATIVE_DISCUSSION} (which is devoted to the averaging over the effective 
detected energy band and to the collisional decoherence) and in Sec.~\ref{AVERAGING_OVER_POSITION} (which discusses
the averaging over the region of neutrino emission inside the sun). The coherence criteria and some of the numerical
results are summarized in Fig.~\ref{FIGURE_CRITERIA}. In Sec.~\ref{RESULTS} we derive the exact numerical 
results for time variations of the detected neutrino flux. Section~\ref{NEUTRINO_LINE} with 
Figs.~\ref{FIGURE_LINE_ENERGY},~\ref{FIGURE_LINE_MAX_MIN} presents the exact analytical calculations
and numerical results for the case of a solar neutrino line. Section~\ref{CONTINUOUS_SPECTRUM} with
Figs.~\ref{FIGURE_CONTINUOUS_OPTIMAL}--\ref{FIGURE_CONTINUOUS_MAX_MIN} presents the same for the continuous 
neutrino spectrum. For the continuous neutrino spectrum we assume a Gaussian energy response function for 
our detection device. In Sec.~\ref{CONCLUSIONS} we give the conclusions. Finally, in Appendix~A we discuss
a generalization of the concept of neutrino paths, and in Appendix~B we give the formal mathematical proof of 
the averaging out of most of the cross terms between different neutrino paths due to summing over the region of 
neutrino emission. Analytical evaluations of some integrals are given in Appendices~C and~D.


\section{The WKB states}\label{WKB}

Let there be two mass states of the neutrino with masses $m_1$ and $m_2$ (we assume that $m_2>m_1$). 
Then one writes 
\beq
&&|\nu_e\!>\, =\phantom{-}\cos\theta |\nu_1\!> + \sin\theta |\nu_2\!>,
\label{E_NEUTRINO}\\
&&|\nu_x\!>\, = -\sin\theta |\nu_1\!> + \cos\theta |\nu_2\!>,
\label{X_NEUTRINO}
\eeq
where $\nu_e$ refers to the electron flavor and $\nu_x$ to the other flavor, possibly the muon flavor. $\theta$ is 
the vacuum mixing angle (we assume that $0\le\theta<\pi/4$). Further, the electron neutrino flavor interacts 
with the electron density $n_e$ of the sun more strongly than the $|\nu_x\!>$ flavor, so that the $|\nu_1\!>$ and 
$|\nu_2\!>$ states are coupled. (We neglect neutral-current interaction which is the same for both flavors.)
Each of the mass states for a neutrino propagating in the $x$ direction is a two-component state consisting of the 
first and fourth components of the Dirac wavefunction. In an infinite  homogeneous medium, with electron density $n_e$, 
the Dirac equation, in the mass state representation, for the four components of the neutrino wavefunction, 
two for each mass state, is~\cite{P_86,Be_86,B_89,F_62}
\begin{equation}
i \frac{\partial}{\partial t} \left[
\begin{array}{c}
  \bfgamma'_t {\bf U}\\
  \bfgamma'_t {\bf V}  
\end{array} 
\right] = \left\{\left[ 
\begin{array}{cc}
  \bfgamma'_x\,{\hat p}_x + {\bf 1}\,m_1 & {\bf 0}\\ 
   {\bf 0} & \bfgamma'_x\,{\hat p}_x + {\bf 1}\,m_2 
\end{array} 
\right] + \left[ 
\begin{array}{cc}
  {\bf 1}\,g\cos^2\theta & {\bf 1}\,g\cos\theta\sin\theta\\ 
  {\bf 1}\,g\cos\theta\sin\theta & {\bf 1}\,g\sin^2\theta 
\end{array} 
\right] \right\}\left[
\begin{array}{c} 
  {\bf U}\\
  {\bf V}
\end{array} 
\right],
\label{DIRAC_EQUATION}
\end{equation}
where the matrix composed of the elements with $g\define\sqrt{2}G_{\rm F}n_e$ describes neutrino interaction with 
electrons, ${\hat p}_x=-i\,\partial/\partial x$, and we take $\hbar=1$, $c=1$.  In vacuum, when $g=0$, $\bf U$ is 
the two-component wavefunction corresponding to the $m_1$ state and $\bf V$ is the two-component wavefunction 
corresponding to the $m_2$ state. The two-component matrices 
\begin{equation}
\bfgamma'_t =\left[  
\begin{array}{cc}  
  1 & 0  \\ 
  0 & -1  
\end{array}
\right] \hspace{0.5cm}\mbox{ and }\hspace{0.5cm}
\bfgamma'_x = \left[
\begin{array}{cc} 
  0 & 1\\
  -1 & 0  
\end{array}
\right] 
\label{DIRAC_MATRICES}
\end{equation}
are the first and fourth components of the usual Dirac matrices $\bfgamma_t$ and $\bfgamma_x$, while $\bf 1$ is just 
the two by two unit matrix~\cite{F_62}. We treat the neutrino wavefunction as one-dimensional in space.

Now, $g$ depends on $x$ since $n_e$ is nonzero and varies inside the sun, and $g$ is zero between the sun and the 
Earth. (We replace the radial coordinate by $x$ and ignore spherical effects.) Consequently, the solution of the 
Dirac equation is space dependent. We write two-component wavefunctions ${\bf U}(x, t)$ and ${\bf V}(x,t)$ as 
those with the same energy $\cal E$
\begin{equation}
{\bf U}(x, t) = \left[
\begin{array}{c}
  u_1(x, {\cal E})\\
  u_2(x, {\cal E})
\end{array}
\right] 
\exp{\left(-i{\cal E}t\right)}, \hspace{0.5cm}
{\bf V}(x, t) = \left[ 
\begin{array}{c}
  v_1(x, {\cal E})\\
  v_2(x, {\cal E}) 
\end{array}
\right] 
\exp{\left(-i{\cal E}t\right)}.
\label{DIRAC_SPINORS}
\end{equation}

In the limit that $m_1$ and $m_2$ are small compared to ${\cal E}$, the solution of the Dirac 
equation, Eq.~(\ref{DIRAC_EQUATION}), with $\bf U$ and $\bf V$ given by~(\ref{DIRAC_SPINORS})
is 
\beq
\left[
\begin{array}{c}
  u_1\\
  u_2
\end{array}
\right] = {\cal U}(x, {\cal E})
\left[
\begin{array}{c}
  1+m_1/{\cal E}\\
  1
\end{array}
\right]
\hspace{0.5cm}\mbox{and}\hspace{0.5cm}
\left[
\begin{array}{c}
  v_1\\
  v_2
\end{array}
\right] = {\cal V}(x, {\cal E})
\left[
\begin{array}{c}
  1+m_2/{\cal E}\\
  1
\end{array}
\right],
\eeq
where the two spatial scalar functions $\cal U$ and $\cal V$ are determined at energy ${\cal E}$ by
\begin{equation} 
\left[ 
\begin{array}{cc}
  {\hat p}_x-{\cal E}+m_1^2/2{\cal E}+g\cos^2\theta & g\cos\theta\sin\theta \\ 
  g\cos\theta\sin\theta & {\hat p}_x-{\cal E}+m_2^2/2{\cal E}+g\sin^2\theta
\end{array} 
\right] 
\left[
\begin{array}{c} 
  {\cal U}\\
  {\cal V}
\end{array} 
\right] = 0. 
\label{GENERAL_NORNAL_MODE_EQUATION}
\end{equation}

Moreover, we can regard $g$ as a slowly varying space function over most of the distance between the source and 
the Earth with the exception of the spherical resonance region in the sun, where 
$g\approx\Delta\define (m_2^2-m_1^2)/2{\cal E}$. Therefore, we take the space dependence of $\cal U $ and 
$\cal V$ in the WKB approximation~\cite{BO_78} as
\begin{equation}
{\cal U}(x, {\cal E}) = u(x,{\cal E}) \exp{\bigg(i\int\limits^x k\,dx \bigg)}, \hspace{0.5cm}
{\cal V}(x, {\cal E}) = v(x,{\cal E}) \exp{\bigg(i\int\limits^x k\,dx \bigg)},
\label{WKB_APPROXIMATION}
\end{equation}
where $X_{\rm N}$ is the position of neutrino emission, and $u(x,{\cal E})$, $v(x,{\cal E})$ are slowly varying
scalar functions of $x$. As a result, equation~(\ref{GENERAL_NORNAL_MODE_EQUATION}) becomes
\begin{equation} 
\left[ 
\begin{array}{cc}
  k-{\cal E}+m_1^2/2{\cal E}+g\cos^2\theta & g\cos\theta\sin\theta \\ 
  g\cos\theta\sin\theta & k-{\cal E}+m_2^2/2{\cal E}+g\sin^2\theta
\end{array} 
\right] 
\left[
\begin{array}{c} 
  u\\
  v
\end{array} 
\right] = 0. 
\label{NORNAL_MODE_EQUATION}
\end{equation}
The space dependence occurs in $ g(x) $. 

For every $ {\cal E} $ there are two solutions of this normal mode equation. Let 
\begin{equation} 
k={\cal E}-\frac{m_1^2+m_2^2}{4{\cal E}}+k'.
\label{K_PRIME_DEFINITION}
\end{equation}
The determinant of the matrix of normal mode Eq.~(\ref{NORNAL_MODE_EQUATION}) is zero, so $k'$ 
satisfies the equation
\begin{equation} 
{k'}^2+k'g(x)-\frac{1}{4}\Delta^2+\frac{1}{2}\Delta\cdot g(x)\cos{2\theta}=0.
\label{K_PRIME_EQUATION}
\end{equation}
Hence, the two values for the local momenta $k'_\pm$ are 
given by
\begin{equation} 
k'_{\pm}=-\frac{1}{2}g\pm\frac{1}{2}\sqrt{g^2-2\Delta g\cos{2\theta}+\Delta^2}=
-\frac{1}{2}g\pm\frac{1}{2}\sqrt{{(g-\Delta\cos{2\theta})}^2+\Delta^2\sin^2{2\theta}}.
\label{K_PRIME}
\end{equation}
The corresponding solutions for the WKB region are given by the $+$ and $-$ neutrino states,
\beq 
|\psi_{\pm}\!>\phantom{} &=& \left[
\begin{array}{c} 
  u_{\pm}\\ 
  v_{\pm} 
\end{array} 
\right]\exp{\bigg(-i{\cal E}t + i\int\limits^x k_\pm\,dx \bigg)} \nonumber\\
&=& N_\pm\left[
\begin{array}{c} 
  g\sin\theta\cos\theta\\ 
  -k'_{\pm}+\Delta/2-g\cos^2\theta  
\end{array} 
\right]\exp{\bigg(-i{\cal E}t + i\int\limits^x k_\pm\,dx \bigg)},
\label{STATES_FIRST_EQUATION}
\eeq
or alternatively as, 
\begin{equation} 
\left[
\begin{array}{c} 
  u_{\pm}\\ 
  v_{\pm} 
\end{array} 
\right]\exp{\bigg(-i{\cal E}t + i\int\limits^x k_\pm\,dx \bigg)}
=\pm{\tilde N}_\pm\left[ 
\begin{array}{c} 
  k'_{\pm}+\Delta/2+g\sin^2\theta\\  
  -g\sin\theta\cos\theta
\end{array} 
\right]\exp{\bigg(-i{\cal E}t + i\int\limits^x k_\pm\,dx \bigg)},
\label{STATES_SECOND_EQUATION}
\end{equation}
where $N_\pm$ and ${\tilde N}_\pm$ are positive normalization constants. 
Equations~(\ref{K_PRIME_DEFINITION}) and~(\ref{K_PRIME})--(\ref{STATES_SECOND_EQUATION}) give us two
WKB solutions of the wave equation~\cite{Be_86,B_89}. Note that these $+$ and $-$ solutions (states) are 
orthogonal at each point.

At the Earth, $X_{\rm E}$, when $g=0$, we have from Eqs.~(\ref{K_PRIME}),~(\ref{K_PRIME_DEFINITION}) 
\begin{equation}
k'_\pm(X_{\rm E})=\pm\Delta/2,
\label{K_AT_EARTH}
\end{equation}
and
\beq
k_+(X_{\rm E})={\cal E}-m_1^2/2{\cal E}, \qquad
k_-(X_{\rm E})={\cal E}-m_2^2/2{\cal E}.
\eeq
In this case equations~(\ref{STATES_SECOND_EQUATION}),~(\ref{STATES_FIRST_EQUATION}) give us
\begin{equation} 
\left[ 
\begin{array}{c} 
  u_{+}(X_{\rm E}) \\ 
  v_{+}(X_{\rm E}) 
\end{array} 
\right] = \left[ 
\begin{array}{c} 
  1 \\ 
  0 
\end{array} 
\right], \hspace{0.5cm}
\left[ 
\begin{array}{c} 
  u_{-}(X_{\rm E}) \\ 
  v_{-}(X_{\rm E}) 
\end{array} 
\right] = \left[ 
\begin{array}{c} 
  0 \\ 
  1 
\end{array} 
\right].
\label{EARTH}
\end{equation}
In other words, at the Earth
\beq
&&|\nu_+\!>\phantom{}\equiv|\nu_1\!>\phantom{}=\cos\theta|\nu_e\!>\phantom{}-\sin\theta|\nu_x\!>, 
\label{PLUS_AT_EARTH}\\
&&|\nu_-\!>\phantom{}\equiv|\nu_2\!>\phantom{}=\sin\theta|\nu_e\!>\phantom{}+\cos\theta|\nu_x\!>.
\label{MINUS_AT_EARTH}
\eeq

\begin{figure}
\vspace{8.0cm}
\includegraphics{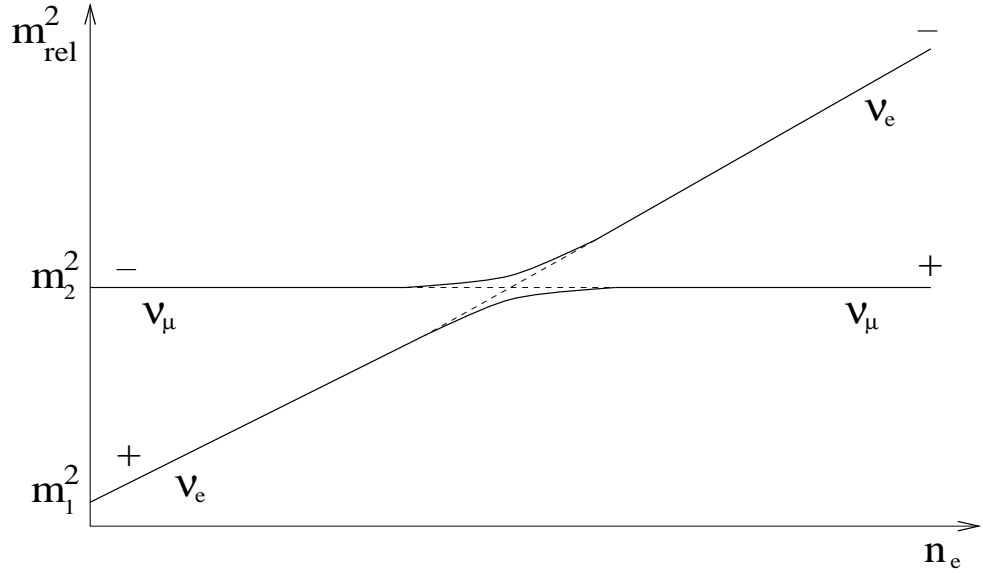}
\caption{The dependence of the neutrino relativistic mass ${\rm m}_{\rm rel}^2={\cal E}^2-k^2$ on the electron density. 
We see that the $-$ state is almost the electron flavor for high densities and it's almost the muon flavor for 
small densities (if the mixing angle is small). The opposite statement is true for the $+$ state. Nuclei emit mostly 
into the $-$ state. If the conversion is adiabatic, the resulting state is the muon difficult to detect at the Earth.}
\label{FIGURE_ADIABATIC_PATHS}
\end{figure}

\begin{figure}
\vspace{12.4cm}
\includegraphics{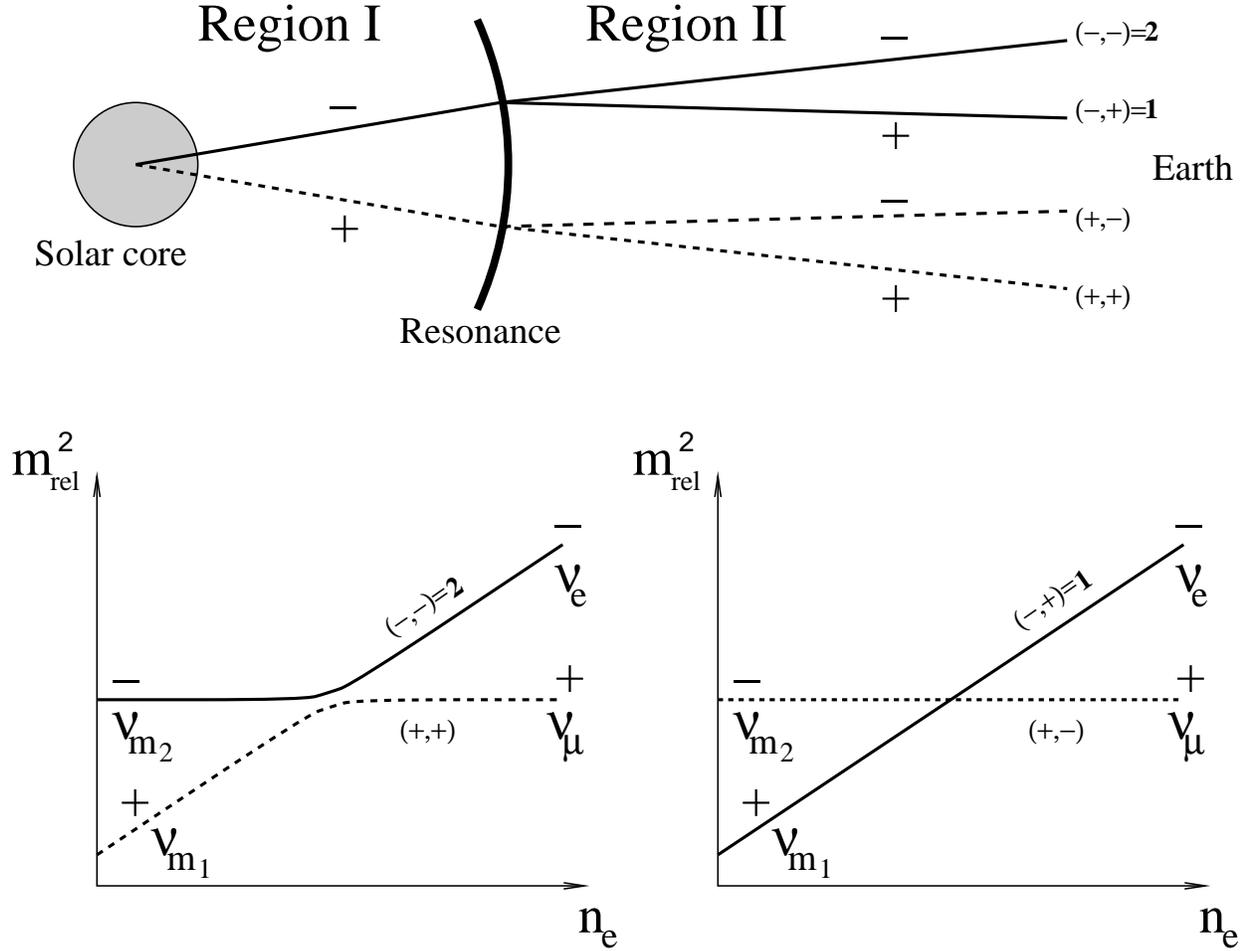}
\caption{All four possible neutrino paths are shown on the upper picture for the case with the resonance 
outside the solar core. 
The left bottom picture shows two adiabatic paths $\Lambda=(-,-)$ and $\Lambda=(+,+)$. This picture can be 
compared with Fig.~\ref{FIGURE_ADIABATIC_PATHS}. On the right bottom picture two nonadiabatic paths 
$\Lambda=(-,+)$ and $\Lambda=(+,-)$ are shown. There are two neutrino eigenfunctions, one indicated by the solid 
lines and the other indicated by the dashed lines.
Each of these eigenfunctions is a liner combination of the corresponding two paths. An electron neutrino emitted 
at the sun can still stay an electron at the Earth due to a nonadiabatic jump from the $-$ to the $+$ state.}
\label{FIGURE_PATHS}
\end{figure}

On the other hand, if we assume that at the position of emission, $X_{\rm N}$, $g\gg\Delta$, then from the same 
Eqs.~(\ref{K_PRIME}),~(\ref{STATES_FIRST_EQUATION}) and~(\ref{STATES_SECOND_EQUATION}) we have
\begin{equation} 
k'_{+}(X_{\rm N})\approx -\frac{1}{2}\Delta\cos{2\theta}, \hspace{0.5cm}
k'_{-}(X_{\rm N})\approx -g+\frac{1}{2}\Delta\cos{2\theta},
\end{equation}
and
\beq
\begin{array}{lcl}
|\nu_+\!>\phantom{} & \approx & (\Delta/g)\sin\theta\cos\theta|\nu_e\!>-
\left[1-(\Delta^2/2g^2)\sin^2\theta\cos^2\theta\right]|\nu_x\!>,\\
|\nu_-\!>\phantom{} & \approx & \left[1-(\Delta^2/2g^2)\sin^2\theta\cos^2\theta\right]|\nu_e\!>\phantom{}+
(\Delta/g)\sin\theta\cos\theta|\nu_x\!>.
\end{array}
\label{PLUS_MINUS_DEEP}
\eeq
In the adiabatic approximation the WKB regions are connected continuously across the resonance region, so 
the $+$ states are connected and the $-$ states are connected. As a result, for the adiabatic theory, in the 
case of $g(X_{\rm N})\gg\Delta$, the emission of the neutrino is primarily into the $-$ state, which at the 
Earth makes a small contribution of $\sin\theta$ (eq.~\ref{MINUS_AT_EARTH}) to the amplitude for an electron 
neutrino detection (if the mixing angle is small), while the contribution into the $+$ state, 
$\approx(\Delta/g)\sin\theta\cos\theta$, is small, but this state makes a finite contribution, 
$\cos\theta$ (eq.~\ref{PLUS_AT_EARTH}) to the amplitude for the neutrino detection. To clarify the statement 
above we plot the dependence of neutrino relativistic mass ${\rm m}_{\rm rel}^2={\cal E}^2-k^2$ on the electron 
density for both states in Fig.~\ref{FIGURE_ADIABATIC_PATHS}~\cite{Be_86}. 
The right side of the figure corresponds to high electron density in the sun interior, the left side to low 
electron density at the Earth.


\section{Nonadiabatic effects and neutrino eigenfunctions}\label{JUMP}

The solution of the Dirac equation, the $+$ and $-$ neutrino states, were found in the previous section under 
the assumption that the WKB approximation is accurate. However, for the resonance region, 
\beq
|g-g_{\rm res}|\simlt\Delta, 
\label{RESONANCE}
\eeq
where 
\beq
g_{\rm res}=\Delta\cos{2\theta},
\label{RESONANCE_G}
\eeq
the WKB approximation breaks down because $k_+$ and $k_-$ are very close at the resonance point,
\beq
k_+-k_-=k'_+-k'_-=\Delta\,\sin{2\theta} \qquad {\rm for}\;\; g=g_{\rm res}
\eeq
[see Eq.~(\ref{K_PRIME})].
As a result, the $+$ state can jump into the $-$ state and vice versa~\cite{P_86,B_89}. One has to find the 
probability of the jump to connect WKB solutions for the neutrino wavefunction in the regions~I and~II.
The jump probability $\delta^2$ can be derived under the assumption that the electron density $n_e$ 
varies linearly throughout the resonance region~\cite{P_86,B_89}
\beq
\delta^2=\exp{\left(-\frac{\pi}{2}\,\Delta\,\frac{\sin^2{2\theta}}{\cos{2\theta}}
{\left[\frac{n_e}{|dn_e/dr|}\right]}_{\rm res}\right)},
\label{JUMP_PROBABILITY_LINEAR}
\eeq
or under the assumption that $n_e$ varies exponentially throughout the resonance region~\cite{KP_88,P_88}
\beq
\delta^2&=&\frac{\exp{\left(2\pi\,r_0\,\Delta\,\cos^2\theta\right)}-1}{\exp{\left(2\pi\,r_0\,\Delta\right)}-1}
=\frac
{\exp{\Big[10^3\cos^2\theta\;{\cal E}_{\rm MeV}^{-1}\:{({m_2^2-m_1^2})}_{-6}\Big]}-1}
{\exp{\Big[10^3\,{\cal E}_{\rm MeV}^{-1}\:{({m_2^2-m_1^2})}_{-6}\Big]}-1}.
\label{JUMP_PROBABILITY}
\eeq
Here $r_0\approx 6.6\times 10^9 {\rm cm}$ is the exponential density scale~\cite{B_89} 
(i.~e.~$n_e(r)=n_e(0)\exp[r/r_0]$), and we use notations ${({m_2^2-m_1^2})}_{-6}=({m_2^2-m_1^2})/10^{-6}\,{\rm eV}^2$ 
and ${\cal E}_{\rm MeV}^{-1}={\cal E}/{\rm MeV}$.
In our numerical calculations we use formula~(\ref{JUMP_PROBABILITY}) for the probability of the jump, because 
it is a good approximation for the whole range of neutrino parameters in the relevant case when the resonance 
electron density is much smaller than the density inside the solar core (see discussion and results below).

Note, that for very small neutrino masses, 
$m_2^2-m_1^2\ll 10^{-9}\,{\rm eV}^2\;{\cal E}_{\rm MeV}$, the jump 
probability~(\ref{JUMP_PROBABILITY}) has the limit $\delta^2\rightarrow \cos^2\theta$. 
It is so-called ``vacuum oscillations'' case, when the neutrino is an oscillating electron neutrino in the vacuum.
This extreme non-adiabatic limit is included into our calculations and results. However, we do not use the term 
``vacuum oscillations'' in our paper, because we believe it could be misleading, and the solar neutrino problem 
should be treated under the MSW theory.

Because of the jump at the resonance point, the $-$ state, emitted in region~I and moving towards the Earth,
becomes a mixture of the $-$ and $+$ solutions in region~II (after passing the resonance), and of course, 
the same statement is true for the $+$ solution. This jump introduces a complication into neutrino propagation
that we describe by the concept of different neutrino paths. It is convenient to introduce a new 
{\it two-dimensional row vector index $\lambda$} to count the two jump possibilities. The index has two possible 
values: $(1,-1)$ and $(1,1)$ corresponding to ``jump'' and ``no jump'' respectively. Then for a neutrino emitted 
into the $-$ state, the formal product of the $-$ and $\lambda$ indices gives two possible values, $(-,+)$ 
and $(-,-)$, which count two possible neutrino paths for this neutrino. For example, the $(-,+)$ path means
that a neutrino originally in the $-$ state (in region~I) is converted into the $+$ state at the resonance 
(and stays $+$ in region~II). The same simple rule is true for the product of the $+$ and $\lambda$ indices 
that counts two possible paths, $(+,-)$ and $(+,+)$, for a neutrino that is emitted into the $+$ state. As a 
result, every neutrino path propagating from the sun to the Earth is defined by its energy $\cal E$, by its 
``sign'' index $\pm$ that indicates the neutrino state at the position of emission (and in region~I), and 
by the row vector index $\lambda$ that indicates whether the neutrino is converted [$\lambda=(1,-1)$] or 
not converted [$\lambda=(1,1)$] from one state to the other at the resonance. We also use a 
{\it two-dimensional row vector index $\Lambda$} that is the product of the indices $\pm$ and $\lambda$ and 
therefore has four possible values, $(-,+)$, $(-,-)$, $(+,-)$ and $(+,+)$ corresponding to four possible neutrino 
paths (see Fig.~\ref{FIGURE_PATHS}). We use the path index $\Lambda$ as well as the $\pm$ and $\lambda$ indices. 
Note, that the $\pm$ index coincides with the first component of $\Lambda$, while $\lambda$ is the formal ratio
of $\Lambda$ to this first component (i.~e.~to the appropriate value of the $\pm$ index).

If a neutrino is emitted into the $+$ state (or into the $-$ state) with amplitude one and the jump probability
at the resonance is $\delta^2$, then the evolution of this neutrino on its way towards the Earth is represented 
by the following equation
\beq
|\psi_\pm(\mbox{region~I})\!>\:&\longrightarrow&
\:\sum\limits_\lambda b_\lambda\: |\psi_{\pm \lambda_2}(\mbox{region~II})\!>,
\label{NONADIABATIC_EIGENFUNCTIONS}
\eeq
where the sum should be taken over two possible values of the index $\lambda$. Here the subscript $\pm\lambda_2$ 
is the product of the $\pm$ index and the second component of the vector index $\lambda$ (i.~e.~$\Lambda_2$ -- the 
second component of $\Lambda$). The absolute values of the amplitudes $b_\lambda$ are easy to derive from the jump 
probability formula~(\ref{JUMP_PROBABILITY})
\beq
\begin{array}{lcl}
|b_{(1,1)}|  &=& {\Bigl(1-\delta^2\Bigr)}^{1/2},\phantom{{\Bigl(1-\delta^2\Bigr)}^{1/2}}\\
|b_{(1,-1)}| &=& \delta.\phantom{{\Bigl(1-\delta^2\Bigr)}^{1/2}}\\
\end{array}
\label{JUMP_AMPLITUDE}
\eeq
It is also obvious that the following equivalence is correct
\beq
\sum\limits_\lambda {|b_\lambda|}^2=1.
\label{FLOW_CONSERVATION}
\eeq
Equation~(\ref{FLOW_CONSERVATION}) simply expresses the neutrino flow conservation law.

As we saw in the previous section, the $+$ and $-$ states are orthogonal solutions of the Dirac equation 
at any point out of the spherical resonance region (i.~e.~in regions~I and~II). An eigenfunction has 
to be an exact solution valid in the resonance region as well. On the other hand, any such solution must be a 
linear combination of the $+$ and $-$ states out of the resonance region (in regions~I and~II). Two orthogonal 
eigenfunction solutions are those represented in Eq.~(\ref{NONADIABATIC_EIGENFUNCTIONS}) [and shown by the solid 
and dashed lines in Fig.~\ref{FIGURE_PATHS}]. It is convenient to take these solutions as our eigenfunctions, since 
they are pure $+$ and $-$ states at the point of neutrino emission.
(We do not include the complex phases in the jump amplitudes ${\tilde b}_\lambda$ and $b_\lambda$ because we 
do not need them. See for references~\cite{L_32,Z_32}.)
As a result, any propagating neutrino with a given energy can be represented as a linear combination of the
two eigenfunctions~(\ref{NONADIABATIC_EIGENFUNCTIONS}), or as a linear combination of four possible paths, $\Lambda$, 
with mathematically only two independent coefficients, all others being given by the jump formulas.

The concept of different neutrino paths and the system of notations using indices $\lambda$ and $\Lambda$ 
can easily be generalized for a case of multiple resonance crossing. The case when the resonance region is inside
the solar core, so there are possibilities of double resonance crossing and no resonance crossing at all, is 
briefly discussed in Appendix~A.


\section{Neutrino emission and detection}\label{EMISSION_DETECTION}

As we have discussed in the introduction section, the correct treatment of nuclear collisions requires a careful
analysis of neutrino emission and detection processes, and proper calculations of neutrino quantum phase, 
emission times and energies. We carry out such analysis and calculations in this section. We make use of
the second quantization method, the method of stationary phase and an expansion in small neutrino masses 
to finally obtain the neutrino wavefunction and the neutrino time dependent flux at the Earth.

\subsection{Second quantization}\label{SECOND_QUANTIZATION}

The charged-current interaction with emission or detection of the neutrinos is given by beta-decay elementary 
particle reactions
\beq
p^++e^-\rlarrows n+\nu_e, \nonumber\\
p^+\rarrow n+e^++\nu_e.
\label{REACTIONS}
\eeq
It is preferable to analyze reactions~(\ref{REACTIONS}) in terms of eigenfunctions of the elementary particles 
in the field of external forces while keeping the interaction Hamiltonian to be zero. These eigenfunctions are 
those for nucleons in a nucleus, eigenfunctions for electrons/positrons in the electric field of the nucleus,
and~(\ref{NONADIABATIC_EIGENFUNCTIONS}) for the neutrinos. 

Let $q_j$ be the amplitude of the $j^{th}$ neutrino oscillator. The index $j$ runs over the two exact neutrino 
eigenfunctions for each energy and over all energy values. (There are two neutrino 
eigenfunctions~(\ref{NONADIABATIC_EIGENFUNCTIONS}) for each energy $\cal E$. We choose them to coincide with 
the $+$ and $-$ states at the position of neutrino emission. As a result, a summation over these eigenfunctions 
is equivalent to a summation over the $\pm$ index.) Under second quantization~\cite{F_32,F_51} the eigenstates 
of the $j^{th}$ oscillator are given by $u_{n_j}(q_j)$, where $n_j=0,1$ represents occupied and unoccupied 
$j^{th}$ state. We imagine these wavefunctions normalized over a large one-dimensional box with a ``volume'' $L$. 
Thus, the second quantized eigenfunction of the neutrinos is
\begin{equation} 
\phi_{n_1, \ldots n_j, \ldots}=u_{n_1}\;\cdots\;u_{n_j}\;\cdots,
\end{equation} 
and it corresponds to $n_j$ neutrinos in the $j^{th}$ state. The general Schr$\ddot{\rm o}$dinger wavefunction is
\begin{equation}
\phi =\sum\limits_{\{n_1, \ldots n_j, \ldots\}} a_{n_1, \ldots n_j, \ldots}\:\phi_{n_1, \ldots n_j, \ldots}.
\end{equation} 

Let us consider a process of electron neutrino emission by a single nucleus, and of electron (or of positron) emission 
if any [see reactions~(\ref{REACTIONS})]. Initially all neutrino states are empty
\beq
\begin{array}{l}
a_{0_1, \ldots 0_j, \ldots}=1, \\
a_{n_1, \ldots n_j, \ldots}=0 \quad {\rm if} \quad \sum\limits_j n_j^2 > 0. 
\end{array}
\label{NEUTRINO_STATES}
\eeq
In our calculations we neglect relativistic effects. If the eigenfunctions of the proton, 
the neutron and the electron are $\psi_p$, $\psi_n$ and $\psi_e$ respectively, the interaction Hamiltonian operator
${\hat{\cal H}}_{int}$ for the creation of an electron neutrino is~\cite{F_51}
\beq
{\hat{\cal H}}_{int}=G_{\rm F} \int {\hat\Psi}_p{\hat\Psi_e}{\hat\Psi}^*_n{\hat\Psi}^*_{\nu_e}\,dL,
\label{INTERACTION_HAMILTONIAN}
\eeq
where the eigenfunctions of Fermi particles are considered as scalar field operators and $G_{\rm F}$ is the Fermi 
constant of interaction. The neutrino wavelength is much larger than the nuclear size, so we use the contact 
interaction approximation. Then the matrix element for the creation of a neutrino into the $j^{th}$ state is
\beq
\left<\Phi_f\,\phi_{0_1, \ldots 1_j, \ldots}\left|{\hat{\cal H}}_{int}\right|\phi_{0_1,\ldots 0_j,\ldots}\,\Phi_i\right>=
G_{\rm F}\,{\left<\psi_j|\nu_e\right>}_{X_{\rm N}}\,
\int\psi_p\psi_e\psi^*_n\,dL=\gamma_{\cal E}\, {\Bigl<\psi_j(X_{\rm N},t)|\nu_e\Bigr>},
\label{MATRIX_ELEMENT}
\eeq
where $\Phi_i$ and $\Phi_f$ are the second quantized wavefunctions for the initial and final states of the proton, 
the neutron and the electron. $\psi_j$ is given by Eqs.~(\ref{NONADIABATIC_EIGENFUNCTIONS}),~(\ref{STATES_FIRST_EQUATION})
and~(\ref{STATES_SECOND_EQUATION}), $X_{\rm N}(t)$ is the position of the nucleus, and we denote the product of all 
factors, which depend on the neutrino energy and are the same for the $+$ and $-$ states, as $\gamma_{\cal E}$.

For an amplitude $a_{0_1, \ldots 1_j, \ldots}$, corresponding to the creation of a neutrino into 
the $j^{th}$ neutrino state, we get
\beq
\frac{d}{dt} a_{0_1, \ldots 1_j, \ldots} &=&
-i\left<\Phi_f\,\phi_{0_1,\ldots 1_j,\ldots}\left|{\hat{\cal H}}_{int}\right|\phi_{0_1,\ldots 0_j,\ldots}\,\Phi_i\right>\;
a_{0_1, \ldots 0_j, \ldots}\, e^{-i{\cal E}_{\!f\:\!\!i\,}t} \nonumber\\
&=& -i\gamma_{\cal E}\, {\Bigl<\psi_j(X_{\rm N},t)|\nu_e\Bigr>}\,e^{-i{\cal E}_{\!f\:\!\!i\,}t}.
\label{NEUTRINO_AMPLITUDE_DERIVATIVE}
\eeq
Here ${\cal E}_{\!f\:\!\!i\,}={\cal E}_i-{\cal E}_f$ is the change of the total energy of the nucleons and the 
electron in the frame co-moving with the nucleus during the neutrino emission, ${\cal E}$ is the energy of the 
neutrino, and we use initial conditions~(\ref{NEUTRINO_STATES}).

For the emitted neutrino we have at the position of emission
\beq
|\psi_\nu\!>=\sum\limits_\pm \int \omega_{\cal E}\:a_{0_1, \ldots 1_j, \ldots}\;|\psi_j\!>\:d{\cal E},
\label{PSI_NU}
\eeq
where 
\beq
\omega_{\cal E}=\frac{L}{2\pi}\,\frac{dk}{d{\cal E}}
\eeq 
is the density of neutrino states, and we use integration over the energy and summation over the $\pm$ index instead of 
the summation over $j$ (remember that our eigenfunctions coincide with the $+$ and $-$ states at the position of 
neutrino emission). 
Integration of Eq.~(\ref{NEUTRINO_AMPLITUDE_DERIVATIVE}) and substitution of the result into 
Eq.~(\ref{PSI_NU}) give us the amplitude of the emitted neutrino
\beq
|\psi_\nu(x,t)\!>=-i\,\sum\limits_\pm\int\!\!\int \gamma_{\cal E}\,\omega_{\cal E}\,
{\Bigl<\psi_j(X_{\rm N}(t'),t')|\nu_e\Bigr>}\;e^{-i{\cal E}_{\!f\:\!\!i\,}t'}\;|\psi_j(x,t)\!>\;dt'\,d{\cal E}.
\label{EMITTED_NEUTRINO}
\eeq
Here $X_{\rm N}(t')$ and $t'$ are the position and the time of neutrino emission. 
The density of states $\omega_{\cal E}$ is approximately the same 
for the $+$ and $-$ states neglecting terms of the order of $\Delta/{\cal E}\ll 1$. Now we use
Eqs.~(\ref{E_NEUTRINO}),~(\ref{STATES_FIRST_EQUATION}) and~(\ref{NONADIABATIC_EIGENFUNCTIONS}) for 
${\left<\psi_j(X_{\rm N},t')|\nu_e\right>}$ and for the eigenfunctions $|\psi_j\!>$ to finally get the 
amplitude of the emitted neutrino at position $X_{\rm E}$ and time $t$ at the Earth
\beq
|\psi_\nu({\rm Earth})\!>=-i\sum\limits_\pm\sum\limits_\lambda\int\!\!\!\int \gamma_{\cal E}\omega_{\cal E}b_\lambda
\Bigl[u({\cal E},X_{\rm N})\cos\theta+v({\cal E},X_{\rm N})\sin\theta\Bigr]
\left[\!
\begin{array}{c} 
  u_{\rm E}\\ 
  v_{\rm E}
\end{array} 
\!\right]
e^{i\Phi_\Lambda}\:dt'\,d{\cal E},
\label{DETECTED_NEUTRINO}
\eeq
where the exponent $\Phi_\Lambda$ is given by
\begin{equation}
\Phi_\Lambda = \int_{X_{\rm N}(t')}^{X_{\rm E}}  k_\Lambda\,dx +{\cal E}(t'-t)-{\cal E}_{\!f\:\!\!i\,} t'.
\label{PHI}
\end{equation}
Here index $\Lambda$, as usually, is the product of $\pm$ and $\lambda$ indices. Note that in 
Eq.~(\ref{DETECTED_NEUTRINO}) $u({\cal E},X_{\rm N})$ and $v({\cal E},X_{\rm N})$ are calculated at the position 
of neutrino emission using Eqs.~(\ref{STATES_FIRST_EQUATION}) and~(\ref{STATES_SECOND_EQUATION}). They depend only 
on the index $\pm$, i.~e.~only on the first component of the neutrino path index $\Lambda$. On the other hand, 
$u_{\rm E}$ and $v_{\rm E}$ are calculated at the Earth, therefore they depend only on the second (the last) 
component of the path index $\Lambda$ and they are given by eq.~(\ref{EARTH}). The exponent $\Phi_\Lambda$ is
different for different neutrino paths. It depends on both components of $\Lambda$. This is because $k(x)$ can 
be either $k_+(x)$ or $k_-(x)$ for different space intervals of the integration in Eq.~(\ref{PHI}).

\subsection{Method of stationary phase}\label{STATIONARY_PHASE}

$\Phi_\Lambda$ is a rapidly varying function of the neutrino emission time $t'$ and the neutrino energy ${\cal E}$, 
so we may evaluate the double integral in Eq.~(\ref{DETECTED_NEUTRINO}) by the method of stationary phase~\cite{BO_78}. 
That is, for each of the four neutrino paths, $\Lambda$, we determine $t'$ and ${\cal E}$ by equations
\beq
&&0=\frac{\partial\Phi_\Lambda}{\partial{\cal E}}=\int_{X_{\rm N}(t')}^{X_{\rm E}}\,
\frac{\partial k_\Lambda}{\partial{\cal E}}\,dx+(t'-t),
\label{CAUSALITY}\\
&&0=\frac{\partial\Phi_\Lambda}{\partial t'}=-k_\Lambda\frac{\partial X_{\rm N}(t')}{\partial t'}+
({\cal E}-{\cal E}_{\!f\:\!\!i\,}).
\label{DOPPLER_SHIFT}
\eeq
Each of the four paths, $\Lambda$, gives a contribution to the neutrino wavefunction at the Earth at detection 
time $t$. These contributions come from four different values of $t'$ and ${\cal E}$ (one pair for each path).
In other words, each neutrino path has different emission time and energy.

Equation~(\ref{CAUSALITY}) essentially yields causality. That is, since $\partial k_\Lambda/\partial{\cal E}$ is 
the reciprocal of the velocity, this equation expresses the fact that the difference between the time of emission 
$t'$ and that of reception $t$ should be the time of flight for the neutrino path between the instantaneous 
position of the nucleus $X_{\rm N}(t')$ at the time of emission and the position $X_{\rm E}$ where the neutrino 
is received. Equation~(\ref{DOPPLER_SHIFT}) simply represents the Doppler shift of the energy of the emitted 
neutrino path due to the motion of the nucleus. Since $k_\Lambda(x)$ is different for neutrino paths (marked by 
different index $\Lambda$, or alternatively by indices $\pm$ and $\lambda$), the times of emission 
$t'$ are different and therefore the nucleus is at a different place $X_{\rm N}(t')$ and has a different velocity 
at these times. We will solve Eqs.~(\ref{CAUSALITY}) and~(\ref{DOPPLER_SHIFT}) for the four solutions for $t'$ 
and $\cal E$ as expansions in the neutrino masses.

To zero order we neglect neutrino masses with respect to the energy. Then all four neutrino paths are emitted at 
the same time and place, $t_0$ and $X_0$, and with the same energy ${\cal E}_0$ (in the laboratory frame), which 
are given by Eqs.~(\ref{CAUSALITY}) and~(\ref{DOPPLER_SHIFT}) to zero order
\beq
X_{\rm E}-X_0=t-t_0, 
\label{X0_T0}
\eeq
\beq
{\cal E}_0=\frac{{\cal E}_{\!f\:\!\!i\,}}{1-V_x}.
\label{E_0}
\eeq
We use these solutions~(\ref{X0_T0}) and~(\ref{E_0}) in the zero neutrino mass approximation to calculate all factors 
in Eq.~(\ref{DETECTED_NEUTRINO}) except the important exponential term $\exp[i\Phi_\Lambda({\cal E}, t')]$, for which 
we still need to solve Eqs.~(\ref{CAUSALITY}) and~(\ref{DOPPLER_SHIFT}) with more accuracy.

An application of the method of stationary phase to the double integral in Eq.~(\ref{DETECTED_NEUTRINO}) 
gives us 
\beq
|\psi_\nu\!>=\Gamma({\cal E}_0)\,\sum\limits_\pm\sum\limits_\lambda b_\lambda\,
\Bigl[u({\cal E}_0,X_0)\cos\theta+v\bigl({\cal E}_0,X_0)\sin\theta\Bigr]
\left[
\begin{array}{c} 
  u_{\rm E}\\ 
  v_{\rm E}
\end{array} 
\right]
\;e^{i\Phi_\Lambda},
\label{NEUTRINO}
\eeq
where $\Gamma({\cal E}_0)$ includes all factors that are essentially the same for all neutrino paths
(i.e.~independent of indices $\pm$ and $\lambda$), and $\Phi_\Lambda$ is given by Eq.~(\ref{PHI}) with $t'$ and 
$\cal E$ obtained by solving Eqs.~(\ref{CAUSALITY}) and~(\ref{DOPPLER_SHIFT}) for each path.

The probability of neutrino detection is proportional to the square of a interaction matrix element that is
similar to the matrix element~(\ref{MATRIX_ELEMENT}). The probability of detection of a neutrino emitted by one 
nucleus is proportional to ${|<\!\nu_e|\psi_\nu\!>|}^2$, where $|\psi_\nu\!>$ is given by eq.~(\ref{NEUTRINO}).
As a result, the neutrino spectrum, ${\cal P}({\cal E}_0)$, that we detect, is
\beq
{\cal P}({\cal E}_0)={\cal P}_{\rm ST}({\cal E}_0)\:
\Bigl\langle{\cal P}_{\rm MSW}({\cal E}_0)\Bigr\rangle,
\label{SPECTRUM}
\eeq
where
\beq
{\cal P}_{\rm MSW}={\biggl|\,\sum\limits_\pm\sum\limits_\lambda b_\lambda\,
\Bigl[u({\cal E}_0,X_0)\cos\theta+v\bigl({\cal E}_0,X_0)\sin\theta\Bigr]
\Bigl[u_{\rm E}\cos\theta+v_{\rm E}\sin\theta\Bigr]
\;e^{i\Phi_\Lambda}\,\biggr|}^{\,2}
\label{MSW_FACTOR_1}
\eeq
is the factor due to the MSW effect and ${\cal P}_{\rm ST}({\cal E}_0)$ is the neutrino energy spectrum that 
we would detect if there were no MSW effect ($\theta=0$, $m_1=m_2=0$, ${\cal P}_{\rm MSW}=1$). 
The brackets $\langle ... \rangle$ in Eq.~(\ref{SPECTRUM}) mean that the MSW factor ${\cal P}_{\rm MSW}$ 
should be averaged over all nuclei, i.e.~over their positions and their motions. In the next sections 
we solve Eqs.~(\ref{CAUSALITY}),~(\ref{DOPPLER_SHIFT}) and we average ${\cal P}_{\rm MSW}$ to obtain 
the detected neutrino spectrum including the MSW effect. Let us rewrite the MSW factor~(\ref{MSW_FACTOR_1}) using 
the path index~$\Lambda$ as
\beq
{\cal P}_{\rm MSW}={\biggl|\,\sum\limits_\Lambda\,B_\Lambda\;e^{i\Phi_\Lambda}\,\biggr|}^{\,2}=
\sum\limits_\Lambda {|B_\Lambda|}^2+\sum\limits_{\Lambda_1}\:\sum\limits_{\Lambda_2\ne\Lambda_1}
|B_{\Lambda_1}B_{\Lambda_2}|\,\cos(\Phi_{\Lambda_2}-\Phi_{\Lambda_1}),
\label{MSW_FACTOR}
\eeq
where
\beq
B_\Lambda\equiv b_\lambda\,\Bigl[u({\cal E}_0,X_0)\cos\theta+v\bigl({\cal E}_0,X_0)\sin\theta\Bigr]
\Bigl[u_{\rm E}\cos\theta+v_{\rm E}\sin\theta\Bigr].
\label{B_LAMBDA}
\eeq
Note that $b_\lambda$ and $B_\Lambda$ are complex numbers. However, we need only the absolute 
values of $B_\Lambda$ as long as we are not interested in phases which are constant in time 
(see Sec.~\ref{RESULTS}). That is why we drop the complex phases ${\rm Arg}(B_{\Lambda_2}B_{\Lambda_1}^*)$
in the cosine in formula~(\ref{MSW_FACTOR}). [We sacrifice mathematical completeness for simplicity
without affecting any final results.]

From Eq.~(\ref{EARTH}) we obtain
\beq
u_{{\rm E},+\,}\cos\theta+v_{{\rm E},+\,}\sin\theta=\cos\theta,    \qquad
u_{{\rm E},-\,}\cos\theta+v_{{\rm E},-\,}\sin\theta=\sin\theta.
\label{PRODUCT_OF_STATES_EARTH}
\eeq
It can also be verified that
\beq
\begin{array}{lcl}
u_+({\cal E}_0,X_0)\cos\theta+v_+\bigl({\cal E}_0,X_0)\sin\theta&=&\cos\theta_{\rm M},\\ 
u_-({\cal E}_0,X_0)\cos\theta+v_-\bigl({\cal E}_0,X_0)\sin\theta&=&\sin\theta_{\rm M},
\end{array}
\label{PRODUCT_OF_STATES_SUN}
\eeq
where the matter mixing angle $\theta_{\rm M}({\cal E}_0,X_0)$ is defined by
\beq
\tan{2\theta_{\rm M}}=\frac{\Delta\sin{2\theta}}{\Delta\cos{2\theta}-g}.
\label{MATTER_THETA}
\eeq
$\theta_{\rm M}$ is equal to $\pi/2$ at a density far above the resonance density ($g\gg\Delta$), to $\pi/4$ at the 
resonance point ($g=\Delta\cos{2\theta}$) and to the vacuum mixing angle $\theta$ in the vacuum ($g=0$)~\cite{B_89}. 
Here as before $\Delta=(m_2^2-m_1^2)/2{\cal E}_0$.

Let us calculate the MSW factor averaged over both the position of emission and the position of detection (i.e.~over 
the annual motion of the Earth). In this case all cross terms (i.e.~the terms that contain phase differences) 
are averaged out and we have only four terms left in Eq.~(\ref{MSW_FACTOR}), corresponding to the four 
possible neutrino paths shown in Fig.~\ref{FIGURE_PATHS}. According to 
Eqs.~(\ref{JUMP_AMPLITUDE}),~(\ref{B_LAMBDA}),~(\ref{PRODUCT_OF_STATES_EARTH}) and~(\ref{PRODUCT_OF_STATES_SUN})
the absolute values of amplitudes of these four terms are
\beq
\begin{array}{l}
|B_{(+,+)}| = {\Bigl(1-\delta^2\Bigr)}^{(1/2)}\cos\theta_{\rm M}\cos\theta,\\
|B_{(+,-)}| = \delta\cos\theta_{\rm M}\sin\theta,\\
|B_{(-,-)}| = {\Bigl(1-\delta^2\Bigr)}^{(1/2)}\sin\theta_{\rm M}\sin\theta,\\
|B_{(-,+)}| = \delta\sin\theta_{\rm M}\cos\theta,
\end{array}
\label{4_B}
\eeq
and the MSW factor reduces to
\beq
{\left\langle{\cal P}_{\rm MSW}\right\rangle}_{X_0,\,t\,}&=&
{\left\langle{{|B_{(+,+)}|}^2+{|B_{(-,-)}|}^2+{|B_{(+,-)}|}^2+{|B_{(-,+)}|}^2}\right\rangle}_{X_0}\nonumber\\
&=&\frac{1}{2}+\left[\frac{1}{2}-\delta^2\right]\cos{2\theta}\,{\left\langle{\cos{2\theta_{\rm M}}}\right\rangle}_{X_0}.
\label{PARKE_FORMULA}
\eeq
This is the Parke's formula~\cite{P_86}.

Without any averaging, the right side of Eq.~(\ref{MSW_FACTOR}) consists of the sum of $10$ different nonzero 
terms, $6$ of which are cross terms, each contains phase difference factor $\cos(\Phi_{\Lambda_2}-\Phi_{\Lambda_1})$. 
These cross terms can potentially result in interference effects.

\subsection{Solutions of the causality and Doppler shift equations}\label{CAUSALTY_DOPPLER}

Let us continue the expansion of  Eqs.~(\ref{CAUSALITY}) and~(\ref{DOPPLER_SHIFT}) as power series in 
$m_1^2/{\cal E}^2\ll 1$, $m_2^2/{\cal E}^2\ll 1$ and $g/{\cal E}\ll 1$, while the nuclear velocity is considered 
finite. We need
\beq
&&t'=t_0+t^{(1)}+t^{(2)}+\cdots,
\label{T_EXPANSION}\\
&&X_{\rm N}=X_0+X^{(1)}+X^{(2)}+\cdots,
\label{X_EXPANSION}\\
&&{\cal E}={\cal E}_0+{\cal E}^{(1)}+{\cal E}^{(2)}+\cdots.
\label{E_EXPANSION}
\eeq
For the motion of the emitting nucleus we assume
\beq
X^{(1)}=V_x t^{(1)}, \qquad X^{(2)}=V_xt^{(2)}+a_x {[t^{(1)}]}^2\Bigr/2,
\label{X_1_X_2}
\eeq
where $V_x$ and $a_x$ are the velocity and acceleration of the nucleus in the $x$ direction at time $t_0$ 
(remember that the $x$ direction points towards the Earth). We have taken the acceleration to be constant, which is
valid in all relative cases (see the next section). Now we substitute~(\ref{T_EXPANSION}) and~(\ref{X_EXPANSION}) 
into the Doppler shift Eq.~(\ref{DOPPLER_SHIFT}) and the causality Eq.~(\ref{CAUSALITY}) and use 
formula~(\ref{K_PRIME_DEFINITION}) for $k$. To zero order, solutions of the causality equation and the 
Doppler shift equation are Eqs.~(\ref{X0_T0}) and~(\ref{E_0}) as before, while the first 
order solutions of the causality and Doppler shift equations allow us to obtain $t^{(1)}$ and ${\cal E}^{(1)}$
\beq
t^{(1)}&=&-\frac{1}{1-V_x}\int_{X_0}^{X_{\rm E}}\,
\left[\frac{m_2^2+m_1^2}{4{\cal E}_0^2}+\frac{\partial k'({\cal E}_0)}{\partial{\cal E}_0}\right]\,dx
\approx -\int_{X_0}^{X_{\rm E}}\,
\left[\frac{m_2^2+m_1^2}{4{\cal E}_0^2}+\frac{\partial k'({\cal E}_0)}{\partial{\cal E}_0}\right]\,dx,
\label{T1}\\
{\cal E}^{(1)}&=&\frac{1}{1-V_x}\,
\left[{\cal E}_0 a_x t^{(1)}+\biggl(-\frac{m_2^2+m_1^2}{4{\cal E}_0}+k'({\cal E}_0,X_0)\biggr)V_x\right]
\approx{\cal E}_0 a_x t^{(1)},
\label{E1}
\eeq
where the last terms are the simplified variants of these two equations under the conditions 
$V_x\ll 1$, $V_x\ll a_x X_{\rm E}$.

We now evaluate $\Phi_\Lambda$ as an expansion in the squared neutrino masses to the second order accuracy. 
First, we use the Taylor expansion for $k({\cal E})$ in $\cal E$ about ${\cal E}_0$ to write Eq.~(\ref{PHI}),
accurate to second order, as 
\beq
\Phi_\Lambda&=&\int_{X_{\rm N}(t')}^{X_{\rm E}} 
\left[k({\cal E}_0)+({\cal E}-{\cal E}_0)\frac{\partial k({\cal E})}{\partial {\cal E}}\right]\,dx
+{\cal E}(t'-t)-{\cal E}_{\!f\:\!\!i\,}t' \nonumber\\ 
&=&\int_{X_{\rm N}(t')}^{X_{\rm E}} 
k({\cal E}_0)\,dx - ({\cal E}-{\cal E}_0)(t'-t) +{\cal E}(t'-t)-{\cal E}_{\!f\:\!\!i\,}t' \nonumber\\
&=&\int_{X_{\rm N}(t')}^{X_{\rm E}}
k({\cal E}_0)\,dx + ({\cal E}_0-{\cal E}_{\!f\:\!\!i\,})(t'-t)-{\cal E}_{\!f\:\!\!i\,} t.
\label{PHI_1}
\eeq
Note that the term that contains the second derivative of $k$ with respect to the energy is of third order.
To obtain the second expression for $\Phi_\Lambda$ above we use Eq.~(\ref{CAUSALITY}). Now using 
expansions~(\ref{T_EXPANSION}) and~(\ref{X_EXPANSION}), the zero order solutions~(\ref{X0_T0}) and~(\ref{E_0}), 
formula~(\ref{K_PRIME_DEFINITION}) and Eqs.~(\ref{X_1_X_2}), we find
\beq
\Phi_\Lambda&=&-{\cal E}_{\!f\:\!\!i\,} t_0+\int_{X_0}^{X_{\rm E}}\,
\left[-\frac{m_2^2+m_1^2}{4{\cal E}_0}+k'({\cal E}_0,x)\right]\,dx \nonumber\\
&&-{\cal E}_0\,\frac{a_x {[t^{(1)}]}^2}{2}
-\left[-\frac{m_2^2+m_1^2}{4{\cal E}_0}+k'({\cal E}_0,X_0)\right]\,V_x t^{(1)} \nonumber\\
&=&-{\cal E}_{\!f\:\!\!i\,} t_0+\int_{X_0}^{X_{\rm E}}\,\left[-\frac{m_2^2+m_1^2}{4{\cal E}_0}+k'({\cal E}_0,x)\right]\,dx-
{\cal E}_0\,\frac{a_x {[t^{(1)}]}^2}{2},
\label{PHI_EXPANSION}
\eeq
where we drop the last term in the first equation because it is of second order and the nuclear velocity is 
small, $V_x\ll 1$. We also neglect the ${(1-V_x)}^{-1}$ factor in Eq.~(\ref{T1}) for $t^{(1)}$. [Note, that we have
also neglected relativistic effects in calculating the process of neutrino emission. These effects give relativistic 
corrections in Eq.~(\ref{NEUTRINO_AMPLITUDE_DERIVATIVE}). They also give correction 
terms to the phase~(\ref{PHI}). These corrections are smaller than the acceleration term $a_x {[t^{(1)}]}^2\Bigl/2$
by the factor $V$.]

In calculating the MSW factor~(\ref{MSW_FACTOR}) we are interested only in differences of phases between neutrino paths. 
Using Eq.~(\ref{PHI_EXPANSION}) we write the phase difference between the two paths $\Lambda_1$ and $\Lambda_2$ as
\beq
\Phi_{\Lambda_2}-\Phi_{\Lambda_1}=\phi_0(X_0,{\cal E}_0)+\phi_a(X_0,{\cal E}_0)a_x,
\label{DELTA_PHI}
\eeq
where
\beq
&&\phi_0=\int_{X_0}^{X_{\rm E}} \left[k'_{\Lambda_2}({\cal E}_0,x)-k'_{\Lambda_1}({\cal E}_0,x)\right]\,dx,
\label{PHI_0}\\
&&\phi_a=-\frac{1}{2}{\cal E}_0
\left(t^{(1)}_{\Lambda_2}-t^{(1)}_{\Lambda_1}\right)
\left(t^{(1)}_{\Lambda_2}+t^{(1)}_{\Lambda_1}\right).
\label{PHI_A}
\eeq
From Eq.~(\ref{T1}) we have
\beq
&&t^{(1)}_{\Lambda_2}-t^{(1)}_{\Lambda_1}=-\int_{X_0}^{X_{\rm E}}
\left[\frac{\partial k'_{\Lambda_2}({\cal E}_0,x)}{\partial{\cal E}_0}-
\frac{\partial k'_{\Lambda_1}({\cal E}_0,x)}{\partial{\cal E}_0}\right]\,dx = 
-\frac{\partial\phi_0}{\partial{\cal E}_0}\, ,
\label{DELTA_TIME}\\
&&t^{(1)}_{\Lambda_2}+t^{(1)}_{\Lambda_1}=-\int_{X_0}^{X_{\rm E}}
\left[\frac{m_2^2+m_1^2}{2{\cal E}_0^2}+\frac{\partial k'_{\Lambda_2}({\cal E}_0,x)}{\partial{\cal E}_0}+
\frac{\partial k'_{\Lambda_1}({\cal E}_0,x)}{\partial{\cal E}_0}\right]\,dx.
\label{SUM_TIME}
\eeq
$\phi_0$ is well-known large phase sensitively dependent on the energy, which can lead to the averaging out of the 
neutrino flux time variations if the detected energy band is not sufficiently small. $\phi_a a_x$ is stochastic 
term due to nuclear collisions, and it can totally average out the flux variations independently of everything else.
Note, that there are two effects of the nuclear acceleration, which produce the additional phase difference 
$\phi_a(X_0)a_x$ in Eq.~(\ref{DELTA_PHI}). First, the additional distance $X^{(2)}$ that the emitting nucleus 
moves because of its instantaneous acceleration, and second, the Doppler shift of neutrino energy ${\cal E}^{(1)}$
because the nucleus has different velocities at different emission times. These two effects give contributions, 
which differ by a factor of two.

Let us consider the case when the indices $\Lambda_1$ and $\Lambda_2$ have different second (last) components, 
i.e.~the case when in the region~II, after passing the resonance, the $\Lambda_1$ neutrino path is different from 
the $\Lambda_2$ neutrino path. Then the main contribution to the integrals in Eqs.~(\ref{DELTA_TIME}) 
and~(\ref{SUM_TIME}) is given by the interval of integration in the vacuum (even for large values of $g$ inside 
the sun, $g\gg\Delta$). As a result, using Eq.~(\ref{K_AT_EARTH}) for $k'$, we have
\beq
&&|t^{(1)}_{\Lambda_2}\pm t^{(1)}_{\Lambda_1}| = \frac{m_2^2\pm m_1^2}{2{\cal E}_0^2}X_{\rm E},
\label{PM_T_1}\\
&&\phi_0\approx X_{\rm E}\:\Delta\approx 3.8\times 10^5
\:\Biggl(\frac{\rm MeV}{{\cal E}_0}\Biggr)\:\Biggl(\frac{m_2^2-m_1^2}{10^{-6}\,{\rm eV}^2}\Biggr),
\label{PHI_0_VACUUM}\\
&&\phi_a = -{\cal E}_0\,X_{\rm E}^2\frac{m_2^4-m_1^4}{8{\cal E}_0^4},
\label{PHI_A_VACUUM}
\eeq
where we drop $X_0$, which is much smaller than $X_{\rm E}$. For the 
ranges of energy and neutrino masses that are interesting for us, equations~(\ref{PM_T_1}) and~(\ref{PHI_A_VACUUM}) 
are sufficiently accurate. However, the phase $\phi_0$ given by eq.~(\ref{PHI_0_VACUUM}) can be very large compared 
to one (and it is independent of accelerations and emission positions). Therefore, we need to calculate the correction 
to $\phi_0$ due to the integration interval in the sun in eq.~(\ref{PHI_0}).

Equations~(\ref{MSW_FACTOR}),~(\ref{B_LAMBDA}) and~(\ref{DELTA_PHI})--(\ref{PHI_A_VACUUM}) give the solution for 
the MSW factor for any given detection time and single emitting nucleus. However, to find the actual detected 
neutrino flux we still have to average the energy spectrum~(\ref{SPECTRUM}) over accelerations and positions of 
the emitting nuclei and to integrate the spectrum over the effective detected energy band.


\section{Averaging the time dependent neutrino flux: qualitative description}\label{QUALITATIVE_DISCUSSION}

Before we present the exact calculations let us consider qualitatively all important physical effects that can average 
out cross terms and kill any observable interference time variations of the neutrino flux. There are 
three effects that can potentially wash out cross terms: the averaging over the effective detected energy band, 
the averaging over nuclear accelerations (over collisions) and the averaging over the neutrino emission region 
inside the sun. 

Let us consider the interference between two different $\Lambda_1$ and $\Lambda_2$ neutrino paths. It is convenient to 
introduce the following two energy parameters. Let
\beq
&&\sigma_t\define\hbar\left/{|t^{(1)}_{\Lambda_2}-t^{(1)}_{\Lambda_1}|}\right.\approx
\frac{\hbar c}{X_{\rm E}}\:\frac{2{\cal E}_0^2}{m_2^2-m_1^2}\approx 
2.6\times 10^{-3}\:{\rm KeV}\;{\Biggl(\frac{{\cal E}_0}{\rm MeV}\Biggr)}^2\:
{\Biggl(\frac{10^{-6}\,{\rm eV}^2}{m_2^2-m_1^2}\Biggr)},
\label{SIGMA_T}\\
&&\sigma_a\define{\cal E}_0a_{\rm H}|t^{(1)}_{\Lambda_2}-t^{(1)}_{\Lambda_1}|\approx
{\cal E}_0\:\frac{a_{\rm H}\,X_{\rm E}}{c^2}\:\frac{m_2^2-m_1^2}{2{\cal E}_0^2}\approx
0.40\:{\rm KeV}\:\Biggl(\frac{\rm MeV}{{\cal E}_0}\Biggr)\:\Biggl(\frac{m_2^2-m_1^2}{10^{-6}\,{\rm eV}^2}\Biggr).
\label{SIGMA_A}
\eeq
Here $a_{\rm H}$ is the typical (Holtsmark normal~\cite{Ch_43}) acceleration of nuclei defined in Eq.~(\ref{A_H}). 
Note that all numerical estimates in this section are done for the case when the $\Lambda_1$ and $\Lambda_2$ paths are 
different in the region~II (and consequently in the vacuum), so we can use Eqs.~(\ref{PM_T_1}) and~(\ref{PHI_A_VACUUM}). 
[In the next section we will see that only in the case when the $\Lambda_1$ and $\Lambda_2$ paths are the same inside 
the sun and are different outside does the cross term between these two paths survive after the averaging over the 
region of the neutrino emission.] The physical meaning of the parameters $\sigma_t$ and $\sigma_a$ is explained by 
the equations
\beq
\phantom{\mbox{\Huge$\int$}}\sigma_t\:\left|\frac{\partial\phi_0}{\partial{\cal E}_0}\right|&=&1,
\label{SIGMA_T_EXPLANATION}\\
\phantom{\mbox{\Huge$\int$}}\sigma_a&=&|{\cal E}^{(1)}_{\Lambda_2}-{\cal E}^{(1)}_{\Lambda_1}|,
\label{SIGMA_A_EXPLANATION}\\
\phantom{\mbox{\Huge$\int$}}\sigma_a/\sigma_t&=&
{\cal E}_0\:a_{\rm H}{|t^{(1)}_{\Lambda_2}-t^{(1)}_{\Lambda_1}|}^2 \le
{\cal E}_0\:a_{\rm H}\Bigl|{(t^{(1)}_{\Lambda_2})}^2-{(t^{(1)}_{\Lambda_1})}^2\Bigr|=2|\phi_a|\,a_{\rm H}.
\label{ACCELERATION_CRITERION_1}
\eeq
To obtain Eq.~(\ref{SIGMA_T_EXPLANATION}) we differentiate formula~(\ref{PHI_0}) with respect to the 
energy ${\cal E}_0$ and use Eq.~(\ref{DELTA_TIME}). Equation~(\ref{SIGMA_A_EXPLANATION}) follows directly from 
formula~(\ref{E1}).

First, note that the phase difference $\phi_0$ varies significantly with $\cal E$ over the energy interval 
$\sigma_t$, so, if we integrate the neutrino spectrum over an energy band that is larger than $\sigma_t$, 
the cross term between the $\Lambda_1$ and $\Lambda_2$ paths is averaged out. 
Second, if $\sigma_a/\sigma_t$ is large, then $|\phi_a|\,a_{\rm H}$ is large and the phase 
difference~(\ref{DELTA_PHI}) has a large random component due to the distribution of nuclear 
accelerations, $\bf a$, in magnitude and direction. In this case the cross term is also averaged out.

Let our effective energy band width for neutrino detection be $\sigma$. It can be either the energy 
width of the spectral response function of our detection device (as in the case of the continuous neutrino 
spectrum) or it can be the thermal width of a neutrino line. We thus have the following two qualitative criteria 
for the cross term not to be averaged out
\beq
&\sigma\simlt\sigma_t	\qquad &\mbox{or}\qquad \sigma\simlt
2.6\times 10^{-3}\:{\rm KeV}\;{\Biggl(\frac{{\cal E}_0}{\rm MeV}\Biggr)}^2\:
{\Biggl(\frac{10^{-6}\,{\rm eV}^2}{m_2^2-m_1^2}\Biggr)},
\label{ENERGY_CRITERION}\\
&\sigma_a\simlt\sigma_t	\qquad &\mbox{or}\qquad 
150\:{\Biggl(\frac{\rm MeV}{{\cal E}_0}\Biggr)}^3\:{\Biggl(\frac{m_2^2-m_1^2}{10^{-6}\,{\rm eV}^2}\Biggr)}^2\simlt 1.
\label{ACCELERATION_CRITERION}
\eeq
Here we use formulas~(\ref{SIGMA_T}) and~(\ref{SIGMA_A}) for numerical estimates. 
The second criterion has another simple physical explanation: using 
Eq.~(\ref{ACCELERATION_CRITERION_1}), we rewrite the condition $\sigma_a\simlt\sigma_t$ as
\beq
a_{\rm H} \Bigl|{[t^{(1)}_{\Lambda_2}]}^2-{[t^{(1)}_{\Lambda_1}]}^2\Bigr|\simlt \lambdabar, 
\label{DBW}
\eeq
where $\lambdabar=1/{\cal E}$ is the neutrino De Broglie wavelength. We see that the additional distance that 
the emitting nucleus moves because of its instantaneous acceleration must be less than the neutrino De Broglie 
wavelength for the cross term not to be averaged out. However, the collisional decoherence is also produced by the
Doppler shift of the neutrino energy (neutrino paths have different energies because the emitting nucleus has 
different velocities at different emission times).

\begin{figure}
\vspace{11.5cm}
\includegraphics{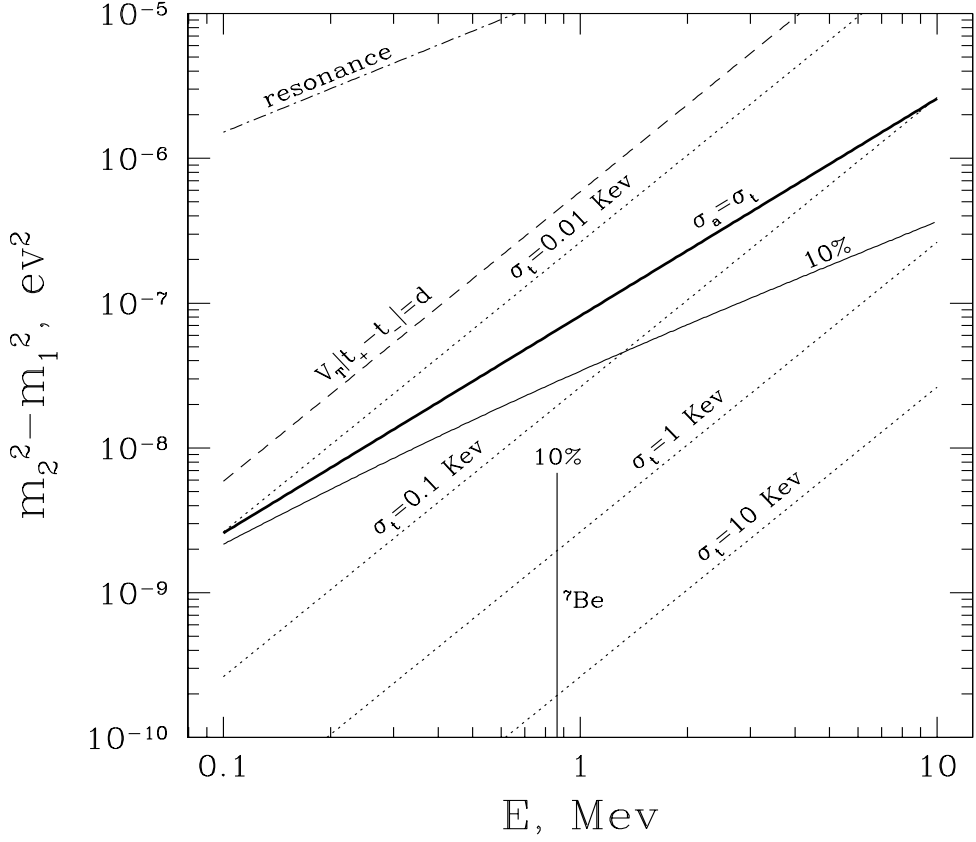}
\caption{A graphical demonstration of the dependence of the qualitative coherence criteria 
(and of some exact results) on two neutrino parameters, the energy and $m_2^2-m_1^2$.
The thick solid line corresponds to the condition $\sigma_a=\sigma_t$. Above this line nuclear collisions 
destroy any coherence, and as a result, there is no time variations of the detected neutrino flux. 
Below this line collisions are unimportant.
The dotted lines correspond to different constant values of $\sigma_t$. The flux time variations are averaged out 
over the effective detected energy band width $\sigma$ for the region of parameter space above the line $\sigma_t=\sigma$.
The vertical and slanted thin solid lines correspond to neutrino flux variations equal to $10\%$ of the standard flux 
(optimized over $\theta$) in two cases respectively: the $0.862\,{\rm MeV}$ $^7$Be solar neutrino line and 
the continuous solar neutrino spectrum. 
The dotted-dashed line shows the extreme case when the resonance electron density is equal to the central solar electron 
density and $\cos{2\theta}\approx 1$. 
The model of constant nuclear acceleration is valid for the region of parameter space below the dashed line 
$V_T|t_+-t_-|=d$ ($V_T$ is the thermal velocity of nuclei and $d$ is the mean
interparticle distance in the region of the solar core).}
\label{FIGURE_CRITERIA}
\end{figure}

Figure~\ref{FIGURE_CRITERIA} illustrates criteria~(\ref{ENERGY_CRITERION}),~(\ref{ACCELERATION_CRITERION}) in 
a convenient form. This figure is plotted for the relevant case, when the $\Lambda_1$ and $\Lambda_2$ neutrino 
paths are different in the vacuum (see the discussion in the next subsection). The thick solid line in the figure 
corresponds to the equality $\sigma_a=\sigma_t$. The criterion~(\ref{ACCELERATION_CRITERION}) is fulfilled in the 
region of parameter space below this line. In other words, above this line collisions destroy any coherence between 
neutrino paths. The dotted lines correspond to different constant values of $\sigma_t$. According to 
criterion~(\ref{ENERGY_CRITERION}), for every given (or chosen) value of the effective detected energy band width 
$\sigma$ we must be in the region below the corresponding dotted line $\sigma_t=\sigma$ for the cross term not to be 
averaged out with the integration of the neutrino spectrum over this energy band. For example, 
for the $0.862\,{\rm MeV}$ $^7$Be solar neutrino line, $\sigma$ is actually the line 
Doppler broadening energy width, which is about $1\,{\rm KeV}$, and we find from Fig.~\ref{FIGURE_CRITERIA} that 
we can observe significant neutrino flux variations only if $m_2^2-m_1^2\simlt 3\times 10^{-9}\,{\rm eV}^2$.

We see from Fig.~\ref{FIGURE_CRITERIA} that unless the effective detected energy band width $\sigma$ 
is smaller than $\sim 0.1\,{\rm KeV}$, the corresponding dotted line $\sigma_t=\sigma$ lies below the thick solid 
line $\sigma_a=\sigma_t$, and therefore collisional coherence criterion~(\ref{ACCELERATION_CRITERION}) is redundant, 
and only criterion~(\ref{ENERGY_CRITERION}) is necessary. Alternatively, above the thick solid line the energy 
averaging criterion $\sigma\simlt\sigma_t$ is redundant. In particular, this is true for the detection of the
$0.862\,{\rm MeV}$ $^7$Be line. 

The vertical thin solid line shows the range of values of $m_2^2-m_1^2$ that correspond to the neutrino flux 
variations of $10\%$ and more of the standard flux (optimized over $\theta$) in the case of the 
$0.862\,{\rm MeV}$ $^7$Be solar neutrino line (we call standard the flux that we would observe if the 
standard theory would be correct, i.e.~if $m_1=m_2=0$). The slanted thin solid line corresponds to $10\%$ 
neutrino flux variations (normalized to the standard flux) for the continuous solar neutrino spectrum and 
for the optimal choice of the detector energy band width and the most favorable value of $\theta$ (see 
Sec.~\ref{CONTINUOUS_SPECTRUM}). Below this line the variations could be larger than $10\%$. Both thin solid 
lines are based on the exact results derived in Sec.~\ref{RESULTS} (for the dependence of these results on the 
mixing angle see Figs.~\ref{FIGURE_LINE_ENERGY}--\ref{FIGURE_CONTINUOUS_MAX_MIN} and the discussion below). Note 
that the $^7$Be vertical line and most of the slanted line lie below the thick solid line, $\sigma_a=\sigma_t$, 
that is in the region where collisional decoherence effects are not significant.

The dotted-dashed line in Fig.~\ref{FIGURE_CRITERIA} shows the maximum possible value of $m_2^2-m_1^2$
for the resonance to exist inside the sun. This line is drawn for the extreme case when the resonance electron 
density $n_{e,\,{\rm res}}$ [see Eq.~(\ref{A_CONDITION_DENSITY})] is equal to the central solar electron 
density and the neutrino mixing angle is small. We see that for the region of parameter space where the collisional 
coherence criterion~(\ref{ACCELERATION_CRITERION}) is satisfied (below the thick solid line), the electron
density inside the neutrino emission region (inside the solar core) is much larger than that at the resonance radius, 
and neutrinos cross the resonance region only once. If we suppose collisional coherence to exist, then 
condition~(\ref{ACCELERATION_CRITERION}) gives us the upper estimate for the resonance electron density
\beq
n_{e,\,{\rm res}}=\frac{(m_2^2-m_1^2)\cos{2\theta}}{2\sqrt{2}G_{\rm F}{\cal E}_0}&\approx&
6.6 N_{\rm A}\cos{2\theta}\:\left(\frac{\rm MeV}{{\cal E}_0}\right) 
\left(\frac{m_2^2-m_1^2}{10^{-6}\,{\rm eV}^2}\right) \nonumber\\
&\simlt& 0.56N_{\rm A}\cos{2\theta}\:{\left(\frac{{\cal E}_0}{\rm MeV}\right)}^{1/2}{\rm cm}^{-3},
\label{A_CONDITION_DENSITY}
\eeq
while the central electron density is about $100N_{\rm A}$.

Finally, the dashed line in Fig.~\ref{FIGURE_CRITERIA} corresponds to the condition 
$V_T|t_+-t_-|=V_T|t^{(1)}_+-t^{(1)}_-|=d$, the inter ion distance. Here $V_T$ is the thermal velocity of nuclei. 
Our approximation of constant nuclear acceleration is valid in the region of parameters that is below the dashed 
line. In our calculations we have considered the emitting nucleus to be accelerated 
uniformly during the time difference $|t_+-t_-|$ with the instantaneous acceleration produced by the nearest 
ion and described by the Holtsmark distribution. In fact, this model of constant nuclear acceleration allows us to 
successfully apply the method of stationary phase and to neglect high order derivatives in the equations of 
nuclear motion. We see that in predicted region of collisional coherence our calculations and results are 
consistent with the assumption of constant nuclear acceleration. This assumption involves only incomplete 
nuclear collisions, which are generally sufficient to decorrelate the neutrino paths.

In Sec.~\ref{RESULTS} we will see how criteria~(\ref{ENERGY_CRITERION}),~(\ref{ACCELERATION_CRITERION}) are 
related to the exact analytical calculations and numerical results for the detected neutrino flux. However, 
we first discuss the important averaging of the neutrino flux over the neutrino emission region inside the sun.

\subsection{Averaging over the position of neutrino emission}\label{AVERAGING_OVER_POSITION}

In this subsection we average the neutrino emission over all the positions of the emitting nuclei. In carrying 
out this average it is possible to consider the more complicated case where the resonance region is inside the 
solar core and the neutrinos can cross the resonance zero, once, or twice, as well as the simpler case where the 
resonance is outside the core and all neutrinos cross it once.

The emission of all neutrinos is concentrated in a region of the solar core, which has the radius 
$R_{\rm core}\approx 0.2 R_\odot$. Because of the high electron density in the core the energy of interaction 
between neutrinos and electrons is strong, and the following estimate is valid:
\beq
g(0)R_{\rm core}>g(R_{\rm core})R_{\rm core}\approx 1500\gg1.
\label{G_R}
\eeq
Let us consider the interference of the $\Lambda_1$ and $\Lambda_2$ neutrino paths. Let these paths be
different in the region of the solar core, i.e.~one of them is the $+$ state and the other is the $-$ state or vice 
versa. Then, according to Eq.~(\ref{K_PRIME}), the difference of their wave numbers inside the core is 
of the order of the larger of $g$ and $m^2$ (so even for very small neutrino masses it is significant).
Now note that the quantum mechanical phase difference $\Phi_{\Lambda_2}-\Phi_{\Lambda_1}$ between the $\Lambda_1$ 
and $\Lambda_2$ neutrino paths is given by Eq.~(\ref{DELTA_PHI}) with the phase $\phi_0$ defined by 
Eq.~(\ref{PHI_0}). The estimate~(\ref{G_R}) shows that the phase $\phi_0$ and, hence, the phase
$\Phi_{\Lambda_2}-\Phi_{\Lambda_1}$ vary strongly with the position of neutrino emission $X_0$.
In this case, the cross term between the $\Lambda_1$ and $\Lambda_2$ neutrino paths given by formula~(\ref{MSW_FACTOR}) 
is the product of the smooth function of $X_0$, ${\cal P}_{\rm ST}|B_{\Lambda_1}B_{\Lambda_2}|$, and the fast 
oscillating function of $X_0$, $\cos[\Phi_{\Lambda_2}-\Phi_{\Lambda_1}]$. The Riemann-Lebesgue lemma~\cite{BO_78} 
guarantees that this cross term is averaged out with the integration over the position of neutrino emission.
(We give the mathematical proof of this statement for the case of a single resonance crossing in Appendix~B).

As a result, the cross term between the $\Lambda_1$ and $\Lambda_2$ neutrino paths is averaged out unless
these paths coincide everywhere in the region of high electron density. Therefore, two important conclusions are 
valid: First, all cross terms are averaged out in the case of a double resonance crossing, when the spherical 
resonance region is in the region of the dense solar core. Because even if the paths only differ beyond the second 
resonance crossing, a significant portion of the last part of the paths is still in the high electron density region.
For the same reason, the single cross term is also averaged out for the case of no resonance crossing at all
(when neutrinos are emitted outside the resonance radius).
Second, for the case of a single resonance crossing, there can be only two cross terms that are not averaged 
out over the region of neutrino emission. They are between the $\Lambda=(+,+)$ and $\Lambda=(+,-)$ paths, and 
between the $\Lambda=(-,-)$ and $\Lambda=(-,+)$ paths. 

Now note, that the cross term between the $(+,-)$ and $(+,+)$ neutrino paths is much smaller than that between the 
$(-,+)$ and $(-,-)$ paths. Indeed, according to Eqs.~(\ref{MSW_FACTOR}),~(\ref{4_B}) the first cross term contains
the factor $\cos^2\theta_{\rm M}$, while the other has the factor $\sin^2\theta_{\rm M}$, and all other factors are 
the same. The collisional coherence criterion can be satisfied only if the electron density inside the neutrino emission 
region is much larger than that at the resonance radius [see the previous section]. 
In this case $g\gg\Delta$ and the matter mixing angle $\theta_{\rm M}$ is very close to $\pi/2$, so that 
$\cos\theta_{\rm M}\approx 0$ and $\sin\theta_{\rm M}\approx 1$ [see Eq.~(\ref{MATTER_THETA})]. In other 
words, neutrinos are emitted mainly in the $-$ state, which is very close to the electron flavor in the region of high 
electron density [see Fig.~\ref{FIGURE_ADIABATIC_PATHS} and also Eqs.~(\ref{PLUS_MINUS_DEEP})].
As a result, only one cross term is potentially important for the neutrino flux variations, namely the cross term 
between the $(-,+)$ and $(-,-)$ neutrino paths. Henceforth, we again consider the case when the solar core is inside 
the region~I and we include only this cross term in Eq.~(\ref{MSW_FACTOR}) for the MSW factor. We can use 
Eqs.~(\ref{PM_T_1}) and~(\ref{PHI_A_VACUUM}) and all other numerical results obtained in the previous section 
for this cross term, because the $(-,+)$ and $(-,-)$ paths are different in the vacuum.


\section{Averaging the time dependent neutrino flux: quantitative results}\label{RESULTS}

In the next two sections we derive the exact analytical formulas and numerical results for the detected neutrino 
flux. We consider two separate cases: that of a neutrino line and that of a continuous neutrino spectrum. 
We have to treat these two cases separately for two reasons. First, the acceleration of the 
emitting nucleus leads to slightly different energies for different neutrino paths [see Eq.~(\ref{E1}]. 
Our detector may have different energy response at these different energy values. 
Second, the characteristic energy band width $\sigma$, of detection, can have different origins. In the first case 
of a neutrino line, $\sigma$ is equal to the line thermal broadening energy width $\sim 1\,{\rm KeV}$, 
and we consider the energy response function of our neutrino detector to be constant throughout the line 
energy profile (i.e.~we do not resolve the line). On the other hand, for the continuous spectrum,
$\sigma$ is a feature of our detector. In this case, to get a definite numerical answer, we choose 
detector's energy response function to be Gaussian with the width $2\sigma$. Of course, in both cases we can 
observe significant time variations of the detected neutrino flux only if the criteria $\sigma_a\simlt\sigma_t$ 
and $\sigma\simlt\sigma_t$ are satisfied. (In addition, for the continuous spectrum we must also have 
$\sigma_a\simlt\sigma$, else the different energies of different neutrino paths will not be simultaneously 
included in detector band width.) These two criteria were qualitatively obtained in 
Sec.~\ref{QUALITATIVE_DISCUSSION} from a simple physical consideration. We now present the accurate calculations
and exact results.

\subsection{The neutrino line spectrum}\label{NEUTRINO_LINE}

Let us first explore the case when the detected portion of the neutrino spectrum is a narrow line, as for the case of 
$^7$Be or ${\rm pep}$ neutrinos. Actually because of the high temperature in a region of
the solar core ($T\approx 1\,{\rm KeV}$), neutrino lines are shifted and broadened in energy~\cite{B_94,B_89}. 
They have non-symmetric energy profiles with characteristic widths approximately equal to $1\,{\rm KeV}$. 

Let us specifically consider, as an example, the $0.862\,{\rm MeV}$ $^7$Be solar neutrino line. Its energy profile 
is non-symmetric and rather complicated~\cite{B_94,B_89}. For the sake of simplicity and for final results to be 
comprehensive, we fit the energy profile by Gaussian function with the half width $\varepsilon=0.67\,{\rm KeV}$. 
In other words, the standard neutrino spectrum (without the MSW effect) is
\beq
{\cal P}_{\rm ST}({\cal E})\propto\exp{\Big[-{({\cal E}-{\cal E}_{\rm c})}^2\Bigl/\varepsilon^2\Big]}.
\label{ENERGY_PROFILE}
\eeq
Here the central energy is ${\cal E}_{\rm c}=862.27\,{\rm KeV}$. We consider the energy 
response function of our neutrino detector to be constant over the whole line energy profile, so 
formulas~(\ref{SPECTRUM})--(\ref{B_LAMBDA}) are valid with the phase difference given by Eq.~(\ref{DELTA_PHI}). 
We first average the neutrino spectrum over nuclear collisions (accelerations), and then over the positions of neutrino 
emission. Finally we integrate the detected spectrum over the energy profile~(\ref{ENERGY_PROFILE}).

Let us first average the MSW factor~(\ref{MSW_FACTOR}) over the distribution of nuclear accelerations~$\bf a$. 
We need to average the cosine of the phase difference between the $\Lambda_1$ and $\Lambda_2$ neutrino paths
\beq
{\left\langle\cos(\Phi_{\Lambda_2}-\Phi_{\Lambda_1})\right\rangle}_{{\bf a}}=
{\left\langle\cos(\phi_0+\phi_a a_x)\right\rangle}_{{\bf a}},
\eeq
where we use Eq.~(\ref{DELTA_PHI}), and $\phi_0$ and $\phi_a$ do not depend on the acceleration. Since the 
direction of the nuclear acceleration ${\bf n}_a$ has a random distribution, we have
\beq
&&{\left\langle\cos(\phi_a a_x)\right\rangle}_{{\bf n}_a}=
\frac{1}{2}\int_0^\pi\,\cos(\phi_a a\cos\vartheta)\sin\vartheta\,d\vartheta=\frac{\sin(\phi_a a)}{\phi_a a},
\\
&&{\left\langle\sin(\phi_a a_x)\right\rangle}_{{\bf n}_a}=
\frac{1}{2}\int_0^\pi\,\sin(\phi_a a\cos\vartheta)\sin\vartheta\,d\vartheta=0.
\eeq
Here $a$ is the absolute value of the acceleration and $\vartheta$ is the angle between the acceleration and 
the $x$ direction, so $a_x=a\cos\vartheta$. As a result,
\beq
{\left\langle\cos(\Phi_{\Lambda_2}-\Phi_{\Lambda_1})\right\rangle}_{{\bf a}}=
\cos(\phi_0)\:{\left\langle\frac{\sin(\phi_a a)}{\phi_a a}\right\rangle}_a.
\label{AVERAGE_V_A}
\eeq

We still must average~(\ref{AVERAGE_V_A}) over the absolute value of the acceleration, $a$.
For this we use the Holtsmark distribution $f_{\rm H}$ of the instantaneous nuclear accelerations~\cite{Ch_43}
\beq
f_{\rm H}(a)=H(\beta)\Bigl/a_{\rm H},
\label{F_H}
\eeq
where $H(\beta)$ is the Holtsmark function and $\beta\define a/a_{\rm H}$. Here
\beq
a_{\rm H}=2\pi{\left(\frac{4}{15}\right)}^{2/3}\frac{Q_{\rm N}}{m_{\rm N}}
{\left\langle Q_i^{3/2}\right\rangle}^{2/3} n_i^{2/3}
\label{A_H}
\eeq
is the {\it normal acceleration}, $Q_{\rm N}$ and $m_{\rm N}$ are the electric charge and the mass of the emitting 
nucleus, ${\left\langle Q_i^{3/2}\right\rangle}$ is the average of the three halves power of the charges of the ions
producing the acceleration, and $n_i$ is the ion density. The Holtsmark function is
\beq
H(\beta)=\frac{2}{\pi\beta}\int\limits_0^\infty x\sin x\,\exp\left[-{(x/\beta)}^{3/2}\right]\,dx.
\label{H}
\eeq
In Appendix~C we show that the average in Eq.~(\ref{AVERAGE_V_A}) reduces to
\beq
{\left\langle\frac{\sin(\phi_a a)}{\phi_a a}\right\rangle}_a=\int_0^\infty 
\frac{\sin(\phi_a a_{\rm H}\beta)}{\phi_a a_{\rm H}\beta}\,
H(\beta)\,d\beta=\exp\left(-{|\phi_a a_{\rm H}|}^{3/2}\right).
\label{ACCELERATION_INTEGRAL}
\eeq

Combining formulas~(\ref{MSW_FACTOR}),~(\ref{AVERAGE_V_A}) and~(\ref{ACCELERATION_INTEGRAL}), we obtain the 
cross term between the $\Lambda_1$ and $\Lambda_2$ paths averaged over nuclear collisions
\beq
{\left\langle |B_{\Lambda_1}B_{\Lambda_2}|\,\cos(\Phi_{\Lambda_2}-\Phi_{\Lambda_1})\right\rangle}_{{\bf a}}=
|B_{\Lambda_1}B_{\Lambda_2}|\,\cos(\phi_0)\:\exp\left(-{|\phi_a a_{\rm H}|}^{3/2}\right).
\label{A_AVERAGE_CROSS_TERM}
\eeq
We see that unless $|\phi_a a_{\rm H}|\simlt 1$, the cross term is averaged out over nuclear 
accelerations in accordance with the results obtained in Sec.~\ref{QUALITATIVE_DISCUSSION} 
[see Eqs.~(\ref{ACCELERATION_CRITERION_1}),~(\ref{ACCELERATION_CRITERION})]. 
From Eq.~(\ref{PHI_A_VACUUM})
\beq
|\phi_a a_{\rm H}|=\frac{{\cal E}_0\,X_{\rm E}}{\hbar c}\;\frac{a_{\rm H}\,X_{\rm E}}{c^2}\;
\frac{m_2^4-m_1^4}{8{\cal E}_0^4}\approx 75\,{\left(\frac{\rm MeV}{{\cal E}_0}\right)}^3\,
\left(\frac{m_2^4-m_1^4}{10^{-12}\,{\rm eV}^4}\right),
\label{A_FACTOR}
\eeq
where $a_{\rm H}$ is calculated at the center of the sun. ($|\phi_a a_{\rm H}|$ is roughly independent of the 
position of neutrino emission.) 

To carry out the average over the position of neutrino emission we first note that only one cross term, between 
the $\Lambda_2=(-,+)$ and $\Lambda_1=(-,-)$ states, is important and that the electron density inside the neutrino 
emission region is much larger than that at the resonance radius (see previous section).
In the integral in Eq.~(\ref{PHI_0}) $k'_{(-,+)}$ and $k'_{(-,-)}$ are different only in region~II. Hence, 
we rewrite the phase $\phi_0$ as
\beq
\phi_0=\int_{X_0}^{X_{\rm E}} \left[k'_{(-,+)}({\cal E}_0,x)-k'_{(-,-)}({\cal E}_0,x)\right]\,dx=
\int_{X_{\rm res}}^{X_{\rm E}}\left[k'_+({\cal E}_0,x)-k'_-({\cal E}_0,x)\right]\,dx.
\label{PHI_0_FINAL}
\eeq
As a result of Eqs.~(\ref{MSW_FACTOR}),~(\ref{A_AVERAGE_CROSS_TERM}), we can write
\beq
\Bigl\langle{\cal P}_{\rm MSW}({\cal E}_0,t)\Bigr\rangle={\langle{\cal P}_{\rm MSW}({\cal E}_0)\rangle}_{t\,}+
A({\cal E}_0)\:\cos{\left[\phi_0({\cal E}_0,t)\right]}.
\label{MSW_FACTOR_FINAL}
\eeq
The MSW factor averaged over time, ${\langle{\cal P}_{\rm MSW}\rangle}_{t\,}$, is given by Parke's 
formula, eq.~(\ref{PARKE_FORMULA}), with $\cos2\theta_{\rm M}=-1$ (since the solar core density is much higher than 
the resonance density)
\beq
{\langle{\cal P}_{\rm MSW}\rangle}_{t\,}=\frac{1}{2}-\left(\,\frac{1}{2}-\delta^2\,\right)\,\cos{2\theta}.
\label{MSW_FACTOR_AVERAGED_FINAL}
\eeq
The amplitude of variations of the MSW factor, which we denote as $A$, is
\beq
A({\cal E}_0)&=&2\:{\left\langle |B_{(-,+)}B_{(-,-)}|\right\rangle}_{X_0}\,
\exp\left(-{|\phi_a a_{\rm H}|}^{3/2}\right) \nonumber\\
&=&\delta{\Bigl(1-\delta^2\Bigr)}^{1/2}\,\sin{2\theta}\:\exp\left(-{|\phi_a a_{\rm H}|}^{3/2}\right).
\label{A}
\eeq
The factor $2$ in this formula arises because we count cross terms in Eq.~(\ref{MSW_FACTOR}) twice.
We use Eqs.~(\ref{4_B}) with $\theta_{\rm M}=\pi/2$ to obtain the final expression for $A$ in
Eq.~(\ref{A}).
Note that it is not necessary to average the phase $\phi_0$ in formula~(\ref{MSW_FACTOR_FINAL})
over the position of neutrino emission. [It is true that $X_{\rm res}$, $\delta$ and the complex phase of 
the factor $B_{(-,+)}B_{(-,-)}^*$, which we dropped in Eq.~(\ref{MSW_FACTOR}), depend 
on the position of neutrino emission $X_0$ because neutrinos emitted at different positions in the region of 
the solar core cross the resonance in slightly different places. However this dependence is negligible for 
the relevant range of parameters where collisional coherence 
conditions~(\ref{ACCELERATION_CRITERION}),~(\ref{A_CONDITION_DENSITY}) are satisfied.
In fact, we see that there is very little dependence of the cross term on the position of emission.]

Now the only thing left to do is the integration over the line 
energy profile. Since the width of the energy profile is narrow, we consider all the factors in 
formula~(\ref{MSW_FACTOR_FINAL}) to be constant over the energy range except the phase $\phi_0({\cal E}_0,t)$. 
The Taylor expansion for this phase about ${\cal E}_{\rm c}$ is
\beq
\phi_0({\cal E}_0,t)&=&\phi_0({\cal E}_{\rm c},t)+\frac{\partial\phi_0}{\partial{\cal E}} ({\cal E}_0-{\cal E}_{\rm c})=
\phi_0({\cal E}_{\rm c},t)-(t^{(1)}_+-t^{(1)}_-)({\cal E}_0-{\cal E}_{\rm c}) \nonumber\\
&=&\phi_0({\cal E}_{\rm c},t)-\frac{\Delta}{{\cal E}_{\rm c}}X_{\rm E}\:({\cal E}_0-{\cal E}_{\rm c}),
\label{PHI_0_ENERGY_EXPANSION}
\eeq
where we use Eqs.~(\ref{PHI_0}),~(\ref{DELTA_TIME}) and~(\ref{PM_T_1}) [the second order Taylor term is never
important]. Then $\cos\phi_0$ averaged over energy profile~(\ref{ENERGY_PROFILE}) is
\beq
{\Bigl\langle\cos{\phi({\cal E}_0)}\Bigr\rangle}_{{\cal E}_0}&=&
\Biggl\{\int_{-\infty}^{\infty} {\cal P}_{\rm ST}({\cal E}_0)\,\cos{[\phi_0({\cal E}_0,t)]}\,d{\cal E}_0\Biggr\}\Biggl/
\Biggl\{\int_{-\infty}^{\infty} {\cal P}_{\rm ST}({\cal E}_0)\,d{\cal E}_0\Biggr\} \nonumber\\
&=&\exp{[-{(X_{\rm E}\Delta\,\varepsilon/2{\cal E}_{\rm c})}^2]}\,
\cos{[\phi_0({\cal E}_{\rm c},t)]}=\exp{[-{(\varepsilon/2\sigma_t)}^2]}\,\cos{[\phi_0({\cal E}_{\rm c},t)]}
\vphantom{\Big[},
\label{PHI_PROFILE_AVERAGED}
\eeq
where $\sigma_t$ is defined by~(\ref{SIGMA_T}) with ${\cal E}_0={\cal E}_{\rm c}$. We see that unless a 
condition~$\varepsilon\simlt\sigma_t$ is satisfied, i.~e.~the line energy width is more narrow than $\sigma_t$, 
the cross term is averaged out over the energy profile. Comparing this result to criterion~(\ref{ENERGY_CRITERION}), 
we see that the characteristic energy band width $\sigma$ for a neutrino line is approximately equal to the Doppler 
broadening width as one expects.

Let $F$ and $F_{\rm ST}$ be the neutrino fluxes detected with and without the MSW effect respectively. Then 
their ratio according to Eqs.~(\ref{SPECTRUM}),~(\ref{MSW_FACTOR_FINAL}) and~(\ref{PHI_PROFILE_AVERAGED}) is
\beq
F\Bigl/F_{\rm ST}={\langle{\cal P}_{\rm MSW}({\cal E}_{\rm c})\rangle}_{t\,}+A({\cal E}_{\rm c})\:
\exp{[-{(\varepsilon/2\sigma_t)}^2]}\:\cos{[\phi_0({\cal E}_{\rm c},t)]},
\label{F_OVER_F_STANDARD_1}
\eeq
where we drop the constant phase $\tan^{-1}{(A_2/A_1)}$.
The eccentricity and the period of the Earth's orbit are $\epsilon=0.0167\ll 1$ and $P=1\,{\rm yr}$ respectively, then 
$X_{\rm E}-{\langle X_{\rm E}\rangle}\approx\epsilon\,{\langle X_{\rm E}\rangle}\cos{(2\pi t/P)}$, where 
${\langle X_{\rm E}\rangle}$ is the mean annual distance between the Earth and the sun. 
The time dependence of $\phi_0({\cal E}_{\rm c},t)$ is easy to obtain from Eq.~(\ref{PHI_0_FINAL})
\beq
\phi_0({\cal E}_{\rm c},t)&=&
\int_{X_{\rm res}}^{{\langle X_{\rm E}\rangle}}\left[k'_+({\cal E}_{\rm c},x)-k'_-({\cal E}_{\rm c},x)\right]\,dx
+\int_{{\langle X_{\rm E}\rangle}}^{X_{\rm E}(t)}\Delta\cdot dx \nonumber\\
&=&{\langle\phi_0\rangle}+\epsilon\:{\langle X_{\rm E}\rangle}\,\Delta\cdot\cos{(2\pi t/P)},
\label{PHI_0_VARIATION}
\eeq
where we use Eq.~(\ref{K_AT_EARTH}) for the difference of $k'_+-k'_-$ in the vacuum, and
\beq
{\langle\phi_0\rangle}=
\int_{X_{\rm res}}^{{\langle X_{\rm E}\rangle}}\left[k'_+({\cal E}_{\rm c},x)-k'_-({\cal E}_{\rm c},x)\right]\,dx.
\label{PHI_0_FINAL_AVERAGE}
\eeq
Then the ratio of the neutrino flux to the standard flux~(\ref{F_OVER_F_STANDARD_1}) finally reduces to
\beq
F\Bigl/F_{\rm ST}={\langle{\cal P}_{\rm MSW}({\cal E}_{\rm c})\rangle}_{t\,}+A_{\rm MSW}({\cal E}_{\rm c})\:
\cos{\left[{\langle\phi_0\rangle}+\zeta\cos{\left(2\pi\frac{t}{P}\right)}\right]},
\label{F_OVER_F_STANDARD}
\eeq
where
\beq
A_{\rm MSW}=\delta{\Bigl(1-\delta^2\Bigr)}^{1/2}\sin{2\theta}\;\exp{\left[-{|\phi_a a_{\rm H}|}^{3/2}\right]}\:
\exp{\Big[-{(\varepsilon/2\sigma_t)}^2\Big]},
\label{A_MSW_LINE}
\eeq
and
\beq
\zeta\define\epsilon\,{\langle X_{\rm E}\rangle}\,\Delta\approx 
6.3\times 10^3\,\left(\frac{\rm MeV}{{\cal E}_0}\right)\,\left(\frac{m_2^2-m_1^2}{10^{-6}\,{\rm eV}^2}\right)\gg 1.
\label{ZETA}
\eeq
We see that provided that $\zeta\ge\pi/2$ the factor $A_{\rm MSW}$ gives the actual value of neutrino flux 
variations normalized to the standard flux, which is about $4.7\times10^9 {\rm cm}^{-2} {\rm s}^{-1}$ for the 
$0.862\,{\rm MeV}$ $^7$Be line. If $\zeta\simlt 1$, 
i.e.~$m_2^2-m_1^2\simlt 1.6\times 10^{-10}\,{\rm eV}^2\;{\cal E}_{\rm MeV}$, then the variations are suppressed 
by a factor $\zeta$ times another factor of order unity that depends on the fractional part of uncertain number 
${\langle\phi_0\rangle}/2\pi$. In this case the phase ${\langle\phi_0\rangle}$ can be treated as a parameter that
must be matched with observations. Particularly, if $\zeta\ll 1$, then $F/F_{\rm ST}$ is a constant in time, which
can differ from the Parke's result!

\begin{figure}
\vspace{7.2cm}
\includegraphics{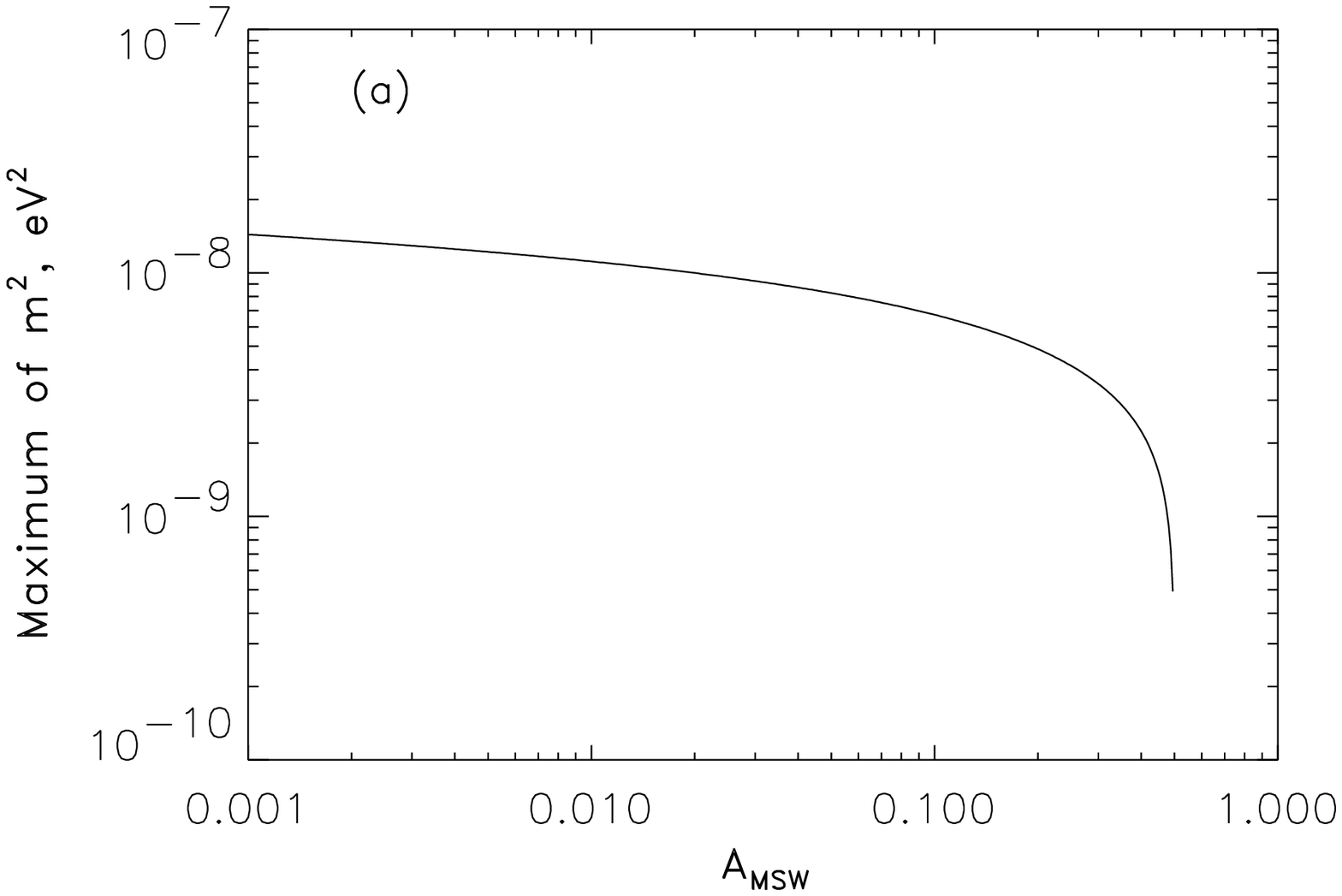}
\includegraphics{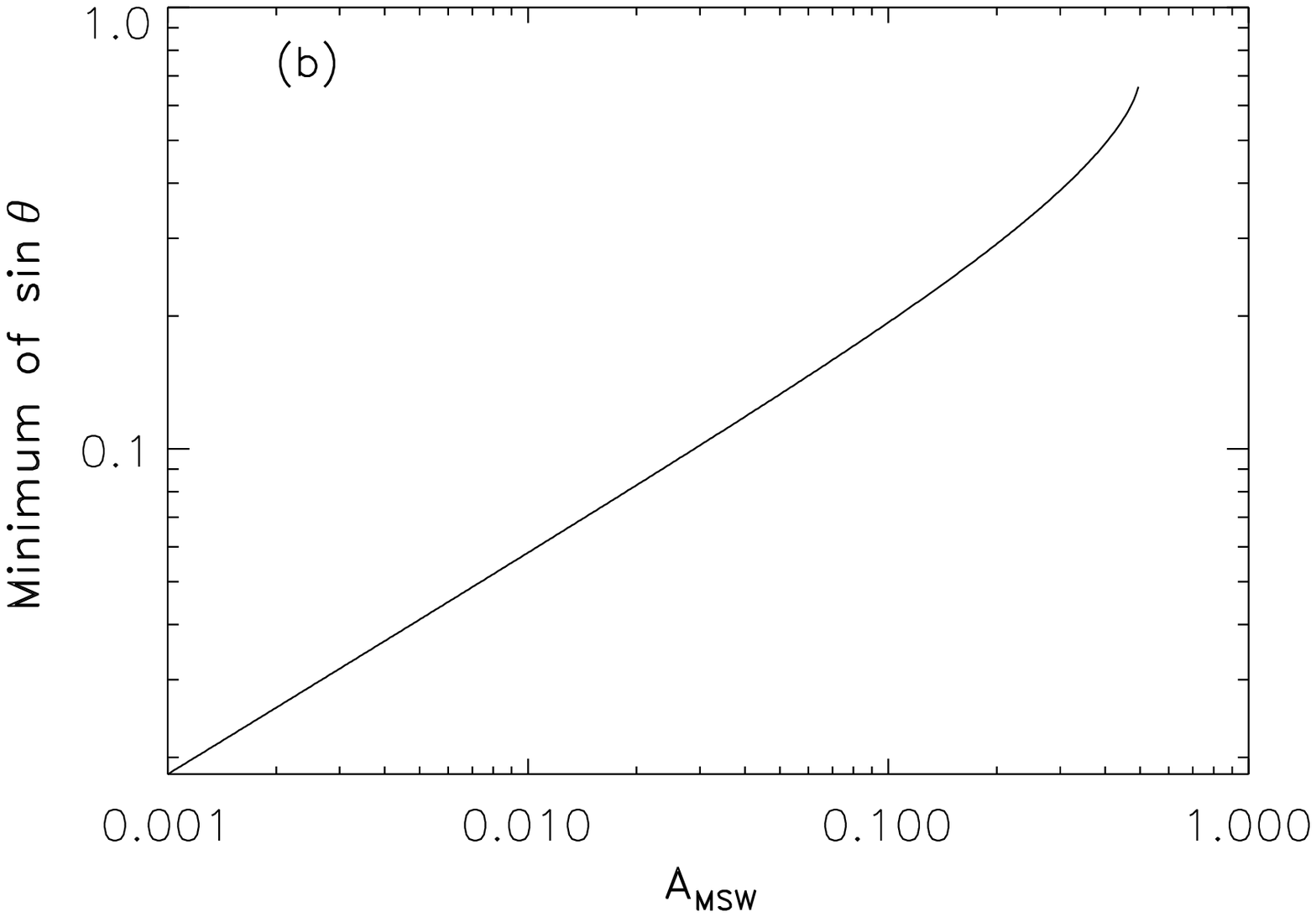}
\caption{
(a):~The maximum possible value of $m^2$ as a function of the desired level of relative 
time variations~$A_{\rm MSW}$ for the $0.862\,{\rm MeV}$ $^7$Be neutrino line.
(b):~The minimum possible value of $\sin\theta$ as a function of~$A_{\rm MSW}$.}
\label{FIGURE_LINE_MAX_MIN}
\end{figure}

How large are the flux variations?
Let us assume that $m_2^2\gg m_1^2$ and use Eq.~(\ref{JUMP_PROBABILITY}) for the jump probability 
and Eq.~(\ref{A_FACTOR}) for the acceleration phase factor $\phi_a a_{\rm H}$. In this case $A_{\rm MSW}$, the 
amplitude of neutrino flux time variations normalized to the standard neutrino flux [see 
Eqs.~(\ref{F_OVER_F_STANDARD}) and~(\ref{A_MSW_LINE})], depends only on two parameters: $m^2\define m_2^2$ 
and the vacuum mixing angle $\theta$. Figure~\ref{FIGURE_LINE_ENERGY} represents contours 
$A_{\rm MSW}={\rm const}$ in the $m^2$ and~$\sin\theta$ parameter space for the $0.862\,{\rm MeV}$ $^7$Be line. 
These contours are based on the exact formula~(\ref{A_MSW_LINE}). 
The maximum possible value of $A_{\rm MSW}$ is $50\%$, which occurs for $\sin{2\theta}\rightarrow 1$ and
the jump probability~$\delta^2=1/2$. We see that there are no significant time variations of the detected 
neutrino flux unless the neutrino mass is quite small, $m^2\sim 5\times 10^{-9}\,{\rm eV}^2$, and the mixing 
angle is quite large, $\sin\theta\sim 0.2$. Figure~\ref{FIGURE_LINE_MAX_MIN}(a) and~\ref{FIGURE_LINE_MAX_MIN}(b) 
support this statement. These two figures show the maximum possible value of $m^2$ and the minimum possible value 
of $\sin\theta$ versus the desired level of the relative (to the standard flux) neutrino flux variations $A_{\rm MSW}$.

If we neglect nuclear acceleration, i.~e.~set $a_{\rm H}$ to be zero in formula~(\ref{A_MSW_LINE}), then all figures
will not noticeably change. This is not surprising, because the acceleration averaging factor in 
formula~(\ref{A_MSW_LINE}), $\exp{[-{|\phi_a a_{\rm H}|}^{3/2}]}$, starts to play its role only when the energy
averaging factor, $\exp{[-{(\varepsilon/2\sigma_t)}^2]}$, is exponentially small. Thus, acceleration averaging is not 
significant for the neutrino line. However, the situation is completely different if we look inside the line (see the
next subsection).

\subsection{The continuous neutrino spectrum}\label{CONTINUOUS_SPECTRUM}

In the case of the detection of the continuous neutrino energy spectrum the detected energy band is determined by 
the energy response function of our detection device and, as we know, must be quite narrow to satisfy 
criterion~(\ref{ENERGY_CRITERION}). Moreover, formulas~(\ref{SPECTRUM})--(\ref{B_LAMBDA}) must be modified 
since the detector energy response function can no longer be considered constant over the observed energy band.
Actually, the detector response is different for neutrino paths with different indices $\Lambda$ because they have
slightly different energies given by Eq.~(\ref{E1}). As a result, we now must allow in our calculations for 
the energy dependence of our detector response.

We consider a neutrino detector with a simple Gaussian energy response function
\beq
S({\cal E})=S_{\rm c}\,\exp{\biggl[-{({\cal E}-{\cal E}_{\rm c})}^2\Bigl/\sigma^2\biggl]},
\label{ENERGY_RESPONSE_FUNCTION}
\eeq
where ${\cal E}_{\rm c}$ is the energy at its center, and $\sigma\ll{\cal E}_{\rm c}$ is its effective energy 
width, which we may choose in order to maximize the time variations of the detected neutrino flux. (Note, that
the variance of $S({\cal E})$ is $\sigma^2/2$.) The detection rate is proportional to the square of the product 
of the electron neutrino amplitude and the square root of the detector response function, so that the observed 
neutrino flux $F$ is  
\beq
F({\cal E}_{\rm c},\sigma)&=&{\cal P}_{\rm ST}({\cal E}_{\rm c})\:\biggl\{\,
\Bigl[\,\sum\limits_\Lambda B_\Lambda^2({\cal E}_{\rm c})\,\Bigr]
\int\limits_{-\infty}^\infty S({\cal E}_0)\,d{\cal E}_0\,+\nonumber\\
&+&\sum\limits_{\Lambda_1}\:\sum\limits_{\Lambda_2\ne\Lambda_1}
|B_{\Lambda_1}({\cal E}_{\rm c})B_{\Lambda_2}({\cal E}_{\rm c})|\,
\Bigl\langle \int\limits_{-\infty}^\infty \sqrt{S({\cal E}_{\Lambda_2})S({\cal E}_{\Lambda_1})}
\cos(\Phi_{\Lambda_2}-\Phi_{\Lambda_1})\,d{\cal E}_0\Bigr\rangle{\vphantom{\bigg|}}_{\!{\bf a},X_{\rm N}}\,\biggr\},
\label{GENERAL_FLUX}
\eeq
where we consider all factors to be constant throughout the detected energy band except the phase difference 
$\Phi_{\Lambda_2}-\Phi_{\Lambda_1}$ and the detector energy response function $S({\cal E})$. [Note that the
energy response function should be calculated at the exact value of neutrino energy ${\cal E}_{\Lambda}$, which is
different from ${\cal E}_0$ by the value~(\ref{E1}).]

Let $\varepsilon\define {\cal E}_0-{\cal E}_{\rm c}\ll{\cal E}_{\rm c}$. Using 
Eqs.~(\ref{E1}),~(\ref{PHI_A}),~(\ref{DELTA_PHI}),~(\ref{PHI_0_ENERGY_EXPANSION}) 
and~(\ref{ENERGY_RESPONSE_FUNCTION}) we have
\beq
&&\Bigl\langle \int\limits_{-\infty}^\infty \sqrt{S({\cal E}_{\Lambda_2})S({\cal E}_{\Lambda_1})}
\cos(\Phi_{\Lambda_2}-\Phi_{\Lambda_1})\,d{\cal E}_0\Bigr\rangle{\vphantom{\bigg|}}_{\!\bf a}=\nonumber\\
&&\qquad=S_{\rm c}\int\limits_{-\infty}^\infty\int\limits_0^\pi\int\limits_0^\infty 
\exp{\biggl\{-\Bigl[{(\varepsilon+{\cal E}_0t^{(1)}_{\Lambda_2}a\cos\vartheta)}^2+
{(\varepsilon+{\cal E}_0t^{(1)}_{\Lambda_1}a\cos\vartheta)}^2\Bigr]\Bigl/2\sigma^2\biggr\}} \nonumber\\
&&\qquad\times\cos{\biggl\{\phi_0({\cal E}_{\rm c})-\varepsilon[t^{(1)}_{\Lambda_2}-t^{(1)}_{\Lambda_1}]-
\frac{1}{2}{\cal E}_0[{(t^{(1)}_{\Lambda_2})}^2-{(t^{(1)}_{\Lambda_1})}^2]\,a\cos\vartheta\biggr\}}f_{\rm H}(a)\,da\:
\frac{1}{2}\sin\vartheta\,d\vartheta\:d\varepsilon.
\eeq
Here again $\vartheta$ is the angle between the nuclear acceleration and the $x$ direction, $a_x=a\cos\vartheta$, 
and $f_{\rm H}(a)$ is the Holtsmark distribution of nuclear accelerations~(\ref{F_H})--(\ref{H}). 
Introducing new variables $\mu\define\cos\vartheta$ and 
$x\define\varepsilon+{\cal E}_0[t^{(1)}_{\Lambda_2}+t^{(1)}_{\Lambda_1}]a\mu/2$, we get
\beq
&&\Bigl\langle \int\limits_{-\infty}^\infty \sqrt{S({\cal E}_{\Lambda_2})S({\cal E}_{\Lambda_1})}
\cos(\Phi_{\Lambda_2}-\Phi_{\Lambda_1})\,d{\cal E}_0\Bigr\rangle{\vphantom{\bigg|}}_{\!\bf a}=
\frac{S_{\rm c}}{2}\int\limits_{-\infty}^\infty\int\limits_{-1}^1\int\limits_0^\infty
\exp{[-\beta^2\sigma_a^2\,\mu^2/4\sigma^2]}\:\exp{[-x^2/\sigma^2]}\nonumber\\
&&\qquad\times\Bigl\{\cos{[\phi_0({\cal E}_{\rm c})]}\cos{[(t^{(1)}_{\Lambda_2}-t^{(1)}_{\Lambda_1})x]}+
\sin{[\phi_0({\cal E}_{\rm c})]}\sin{[(t^{(1)}_{\Lambda_2}-t^{(1)}_{\Lambda_1})x]}\Bigr\}\:f_{\rm H}(a)\,da\:d\mu\:dx,
\eeq
where we use definition~(\ref{SIGMA_A}) for $\sigma_a$, and as before $\beta\define a/a_{\rm H}$.
The integral over $x$ can easily be done (assuming that $\sigma_a$ is independent of $x\ll{\cal E}_{\rm c}$)
to obtain
\beq
&&\Bigl\langle \int\limits_{-\infty}^\infty \sqrt{S({\cal E}_{\Lambda_2})S({\cal E}_{\Lambda_1})}
\cos(\Phi_{\Lambda_2}-\Phi_{\Lambda_1})\,d{\cal E}_0\Bigr\rangle{\vphantom{\bigg|}}_{\!\bf a}=\nonumber\\
&&\qquad=\pi^{1/2}S_{\rm c}\,\sigma\:\exp{[-\sigma^2/4\sigma_t^2]}\:\cos{[\phi_0({\cal E}_{\rm c})]}\:
\int\limits_{0}^1\int\limits_0^\infty\exp{[-\beta^2\mu^2\sigma_a^2/4\sigma^2]}\:H(\beta)\,d\beta\:d\mu.
\label{CROSS_TERM_1}
\eeq
Here we use Eq.~(\ref{F_H}), and $\sigma_t$ is defined by 
Eqs.~(\ref{SIGMA_T}). The integral above with $H(\beta)$ given by Eq.~(\ref{H})
is calculated in Appendix~D. The result is simply
\beq
&&\Bigl\langle \int\limits_{-\infty}^\infty \sqrt{S({\cal E}_{\Lambda_2})S({\cal E}_{\Lambda_1})}
\cos(\Phi_{\Lambda_2}-\Phi_{\Lambda_1})\,d{\cal E}_0\Bigr\rangle{\vphantom{\bigg|}}_{\!\bf a}=\nonumber\\
&&\qquad=4\,S_{\rm c}\,\frac{\sigma^2}{\sigma_a}\:\exp{[-\sigma^2/4\sigma_t^2]}\:\cos{[\phi_0({\cal E}_{\rm c})]}\:
\int\limits_0^\infty x\exp{\left[-x^3-x^4\frac{\sigma^2}{\sigma_a^2}\right]}\:dx.
\label{CROSS_TERM}
\eeq

As before, there is only one important cross term, that is between the $\Lambda=(-,+)$ and $\Lambda=(-,-)$ neutrino paths. 
Combining Eqs.~(\ref{GENERAL_FLUX}) and~(\ref{CROSS_TERM}) we have
for the neutrino flux
\beq
F({\cal E}_{\rm c},\sigma)=\pi^{1/2}S_{\rm c}\,\sigma {\cal P}_{\rm ST}({\cal E}_{\rm c})
{\langle{\cal P}_{\rm MSW}({\cal E}_{\rm c})\rangle}_{t\,}+S_{\rm c}\,A({\cal E}_{\rm c},\sigma)\,
{\cal P}_{\rm ST}({\cal E}_{\rm c})\,\cos{[\phi_0({\cal E}_{\rm c},t)]},
\label{FLUX}
\eeq
where ${\langle{\cal P}_{\rm MSW}\rangle}_{t\,}$ is given by Parke's 
formula~(\ref{MSW_FACTOR_AVERAGED_FINAL}), the phase $\phi_0({\cal E}_{\rm c},t)$ is given by 
Eq.~(\ref{PHI_0_VARIATION}), and 
\beq
A({\cal E}_{\rm c},\sigma)&=&8\,{\left\langle B_{(-,+)}B_{(-,-)}\right\rangle}_{X_0}\:
\frac{\sigma^2}{\sigma_a}\,\exp{[-\sigma^2/4\sigma_t^2]}\,
\int\limits_0^\infty x\exp{\left[-x^3-x^4\frac{\sigma^2}{\sigma_a^2}\right]}\,dx \nonumber\\
&=&4\,\sigma_t\:\delta{\Bigl(1-\delta^2\Bigr)}^{(1/2)}\sin{2\theta}\;Z(\xi,\alpha).
\label{A_MSW_SIGMA}
\eeq
Here we use Eqs.~(\ref{4_B}) and introduce two new dimensionless variables
\beq
\xi&\define&\sigma/\sigma_t
\label{XI}\\
\alpha&\define&\sigma_t/\sigma_a.
\label{ALPHA}
\eeq
The function $Z(\xi,\alpha)$ is
\beq
Z(\xi,\alpha)=\alpha\,\xi^2\:\exp{[-\xi^2/4]}\:
\int\limits_0^\infty x\exp{\left[-x^3-\alpha^2\xi^2x^4\right]}\:dx.
\label{Z}
\eeq

The physical meaning of $\xi$ and $\alpha$ is easy to understand from 
criteria~(\ref{ENERGY_CRITERION}),~(\ref{ACCELERATION_CRITERION}). The effective energy width $\sigma$ is
the feature of our detector. Hence $\xi$ is a free parameter that may be chosen to detect the 
maximum flux variations according to formula~(\ref{FLUX}). If condition $\sigma\simlt\sigma_t$ is not fulfilled, 
the cross term is averaged out over the energy according to criterion~(\ref{ENERGY_CRITERION}). 
On the other hand, if we choose $\sigma$ to be too small, then we detect only a small number of neutrinos in the 
chosen energy band. [The reason for the factor $\sigma^2$ in eq.~(\ref{A_MSW_SIGMA}) is that the fraction of 
accelerations of the emitting nuclei which are small enough for both neutrino paths to have energies inside the 
$\sigma$ band is proportional to $\sigma$.]
Thus, for a given $\alpha$ there is the optimal value of the parameter $\xi=\sigma/\sigma_t$ 
[$\xi_{\rm opt}(\alpha)$] that maximizes the function $Z(\xi,\alpha)$ 
[$Z_{\rm max}(\alpha)=Z(\xi_{\rm opt},\alpha)$], i.e~{\it it maximizes the detected neutrino flux time variations
in SNUs}. Figure~\ref{FIGURE_CONTINUOUS_OPTIMAL} shows this optimal value of $\xi$ (the solid line) and the corresponding 
maximum value of the function $Z(\xi,\alpha)$ (the dotted line) as functions of the parameter $\alpha=\sigma_t/\sigma_a$.

The standard neutrino flux (without the MSW effect) is simply
\beq
F_{\rm ST}({\cal E}_{\rm c},\sigma)=\pi^{1/2}S_{\rm c}\,\sigma\,{\cal P}_{\rm ST}({\cal E}_{\rm c}).
\label{STANDARD_FLUX}
\eeq
Now using Eq.~(\ref{PHI_0_VARIATION}) and definition~(\ref{ZETA}), we obtain the ratio of the observed 
flux~(\ref{FLUX}) to the standard flux for the optimal choice of $\xi$
\beq
F\Bigl/F_{\rm ST}={\langle{\cal P}_{\rm MSW}({\cal E}_{\rm c})\rangle}_{t\,}+A_{\rm MSW}({\cal E}_{\rm c})\:
\cos{\left[{\langle\phi_0\rangle}+\zeta\cos{\left(2\pi\frac{t}{P}\right)}\right]}
\label{RATIO_OF_FLUXES}
\eeq
[again we now see that we do not need the constant complex phase of the factor $B_{(-,+)}B_{(-,-)}^*$]. 
This formula is similar to Eq.~(\ref{F_OVER_F_STANDARD}), but now $A_{\rm MSW}$ is given by
\beq
A_{\rm MSW}=4\pi^{-1/2}\,\delta{\Bigl(1-\delta^2\Bigr)}^{(1/2)}\sin{2\theta}\;
\frac{Z_{\rm max}(\alpha)}{\xi_{\rm opt}(\alpha)}.
\label{A_MSW_CONTINUOUS}
\eeq

Equations~(\ref{FLUX})--(\ref{RATIO_OF_FLUXES}) represent the final result for the case of the continuous neutrino
spectrum. In the limiting case that $\alpha\longrightarrow 0$, i.e.~$\sigma_a\gg\sigma_t$, we have
\beq
\begin{array}{lcl}
\alpha&\longrightarrow& 0,\\
Z(\xi,\alpha)&\longrightarrow&0.451\alpha\xi^2\exp{(-\xi^2/4)},\\
\xi_{\rm opt}(\alpha)&\longrightarrow&2,\\
Z_{\rm max}(\alpha)&\longrightarrow&0.664\:\alpha\longrightarrow 0,
\end{array}
\eeq
and the flux variations are small in accordance with collisional decoherence criterion~(\ref{ACCELERATION_CRITERION}). 
On the other hand, when $\alpha\longrightarrow \infty$, i.e.~$\sigma_a\ll\sigma_t$, we neglect nuclear 
accelerations and we have
\beq
\begin{array}{lcl}
\alpha&\longrightarrow& \infty,\\
Z(\xi,\alpha)&\longrightarrow&0.443\,\xi\exp{(-\xi^2/4)},\\
\xi_{\rm opt}(\alpha)&\longrightarrow&2^{1/2},\\
Z_{\rm max}(\alpha)&\longrightarrow&0.380.
\end{array}
\eeq
Thus, according to formulas~(\ref{RATIO_OF_FLUXES}) and~(\ref{A_MSW_CONTINUOUS}), the maximum amplitude of the 
neutrino flux variations, divided by the standard neutrino flux passing through the corresponding optimized
energy band width (we optimize it to observe the maximum amplitude of the flux variations in SNU), is $42.8\%$. 
This occurs for $\sin{2\theta}\rightarrow 1$ and the jump probability~$\delta^2=1/2$.
Note, that we chose $\xi$ in order to maximize the relative amplitude of the flux variations, it would be larger.

\begin{figure}
\vspace{14.8cm}
\includegraphics{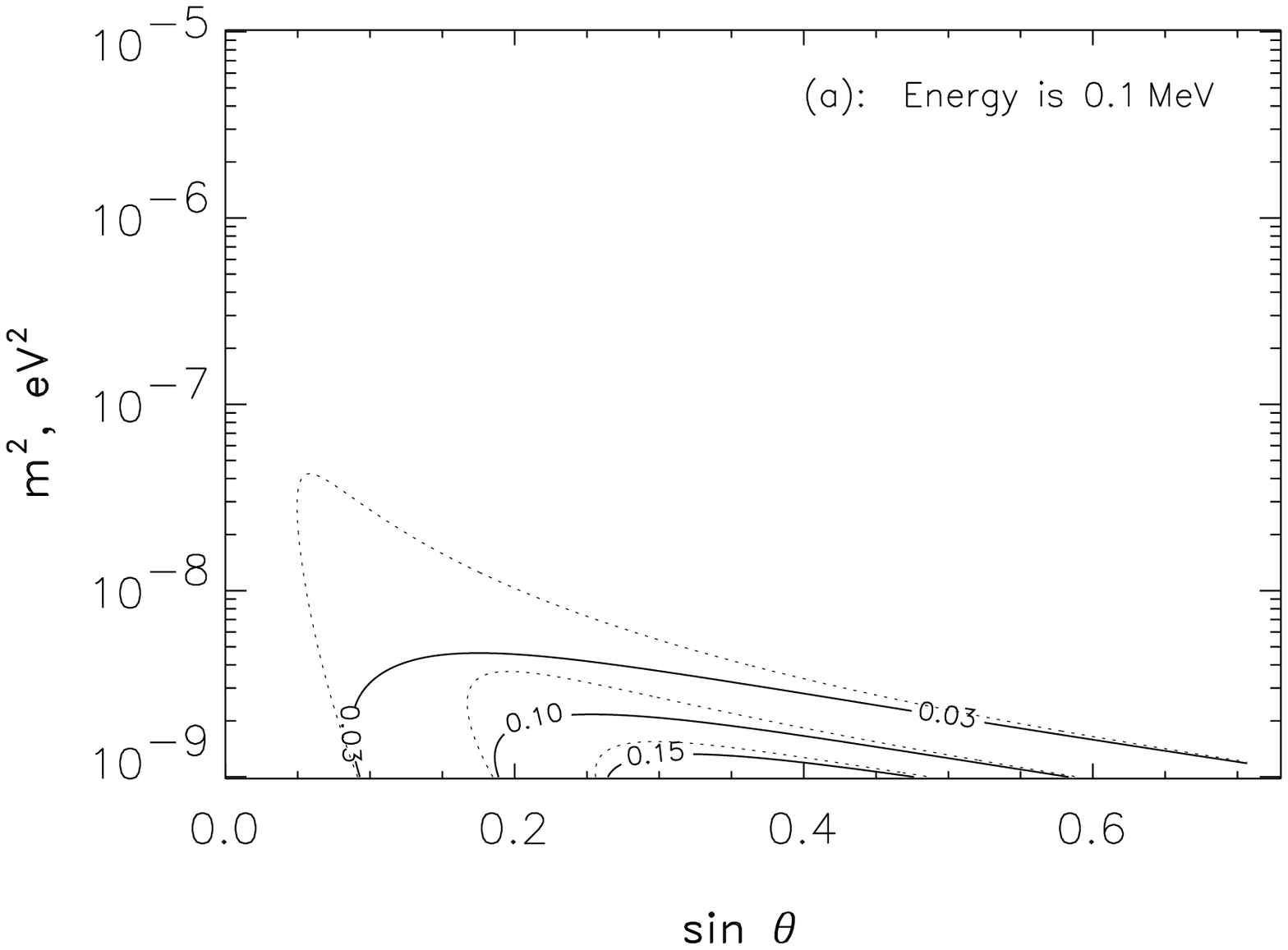}
\includegraphics{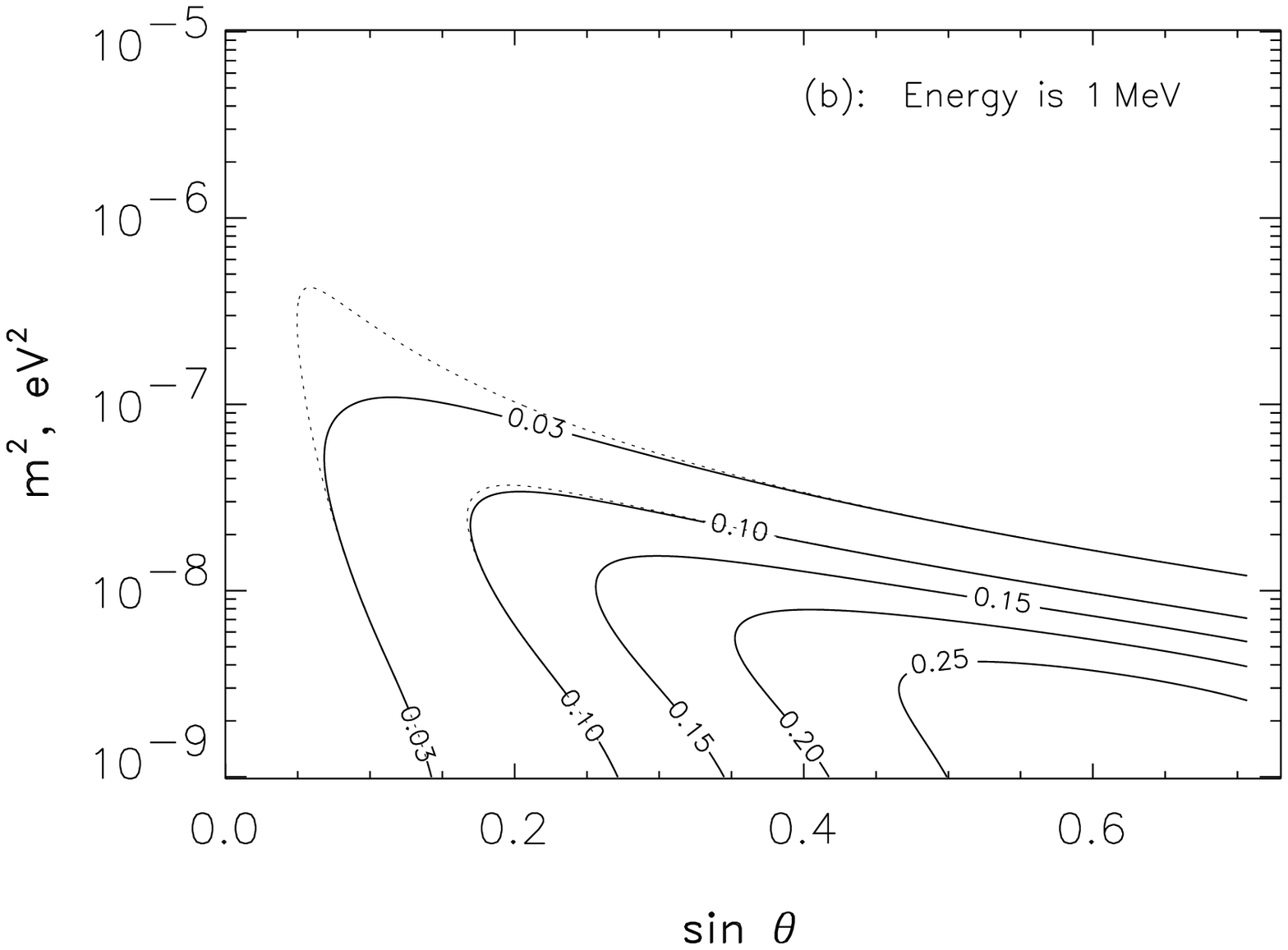}
\includegraphics{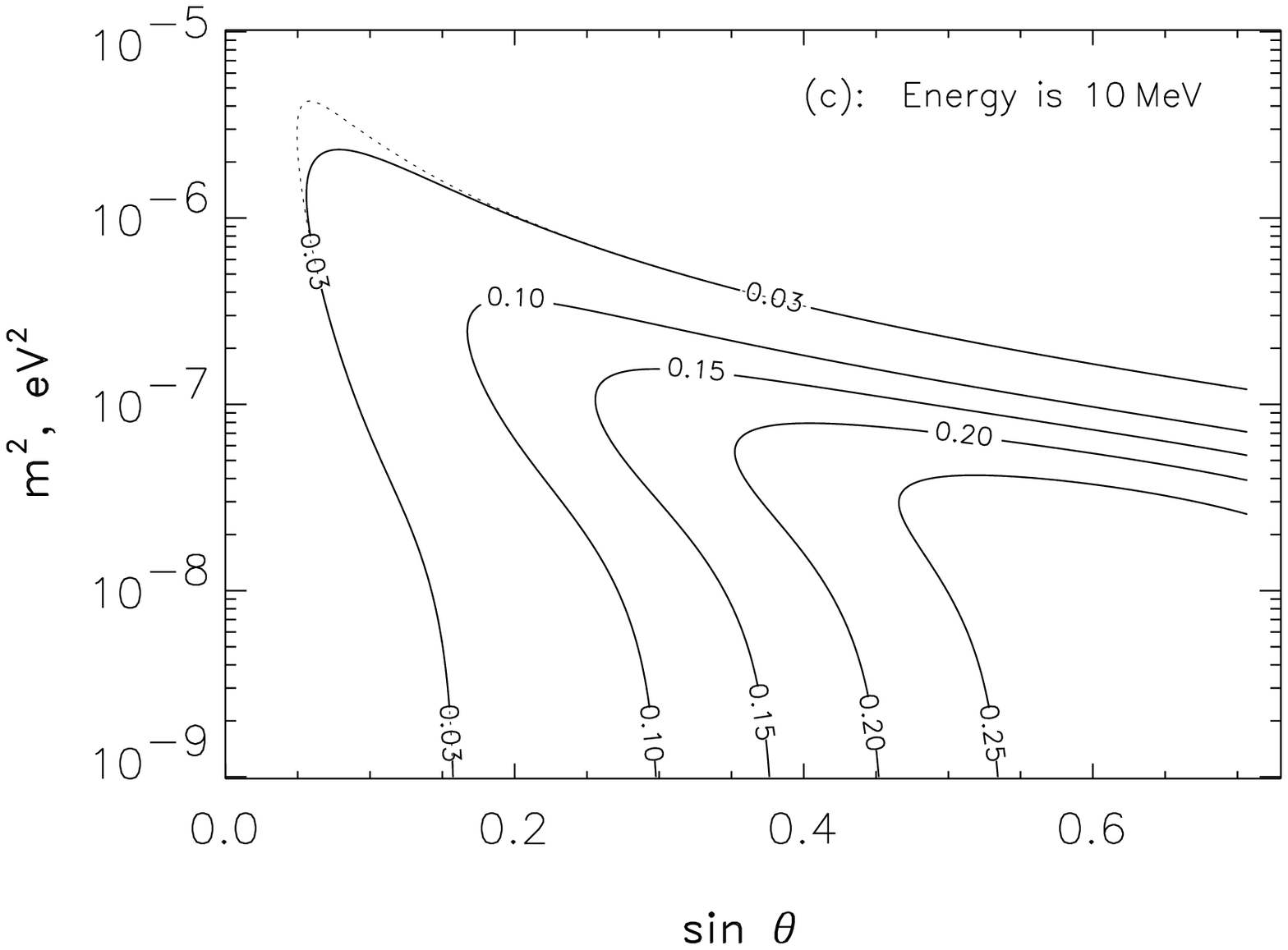}
\includegraphics{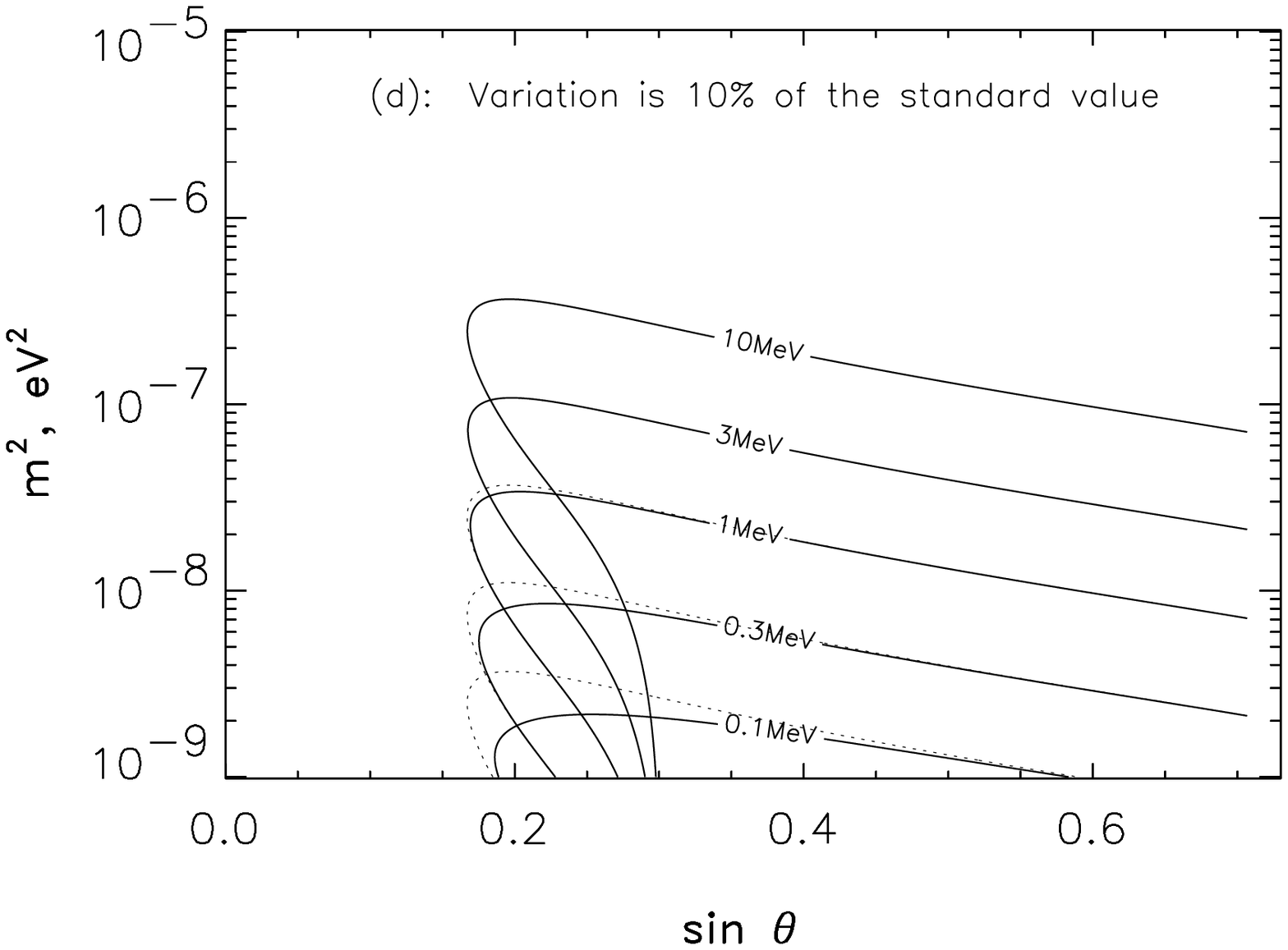}
\caption{
(a),~(b) and~(c):~Contours $A_{\rm MSW}={\rm const}$ are shown for observation of the continuous
neutrino spectrum at different energies (solid lines). These curves represent the 
amplitude of neutrino flux time variations normalized to the standard neutrino flux.
(d):~Constant energy curves ${\cal E}={\rm const}$ are shown for $A_{\rm MSW}=10\%$ variation level (solid lines).
The dotted lines show how the contours would look if we neglect nuclear acceleration.}
\label{FIGURE_CONTINUOUS_ENERGY}
\end{figure}

\begin{figure}
\vspace{8.7cm}
\includegraphics{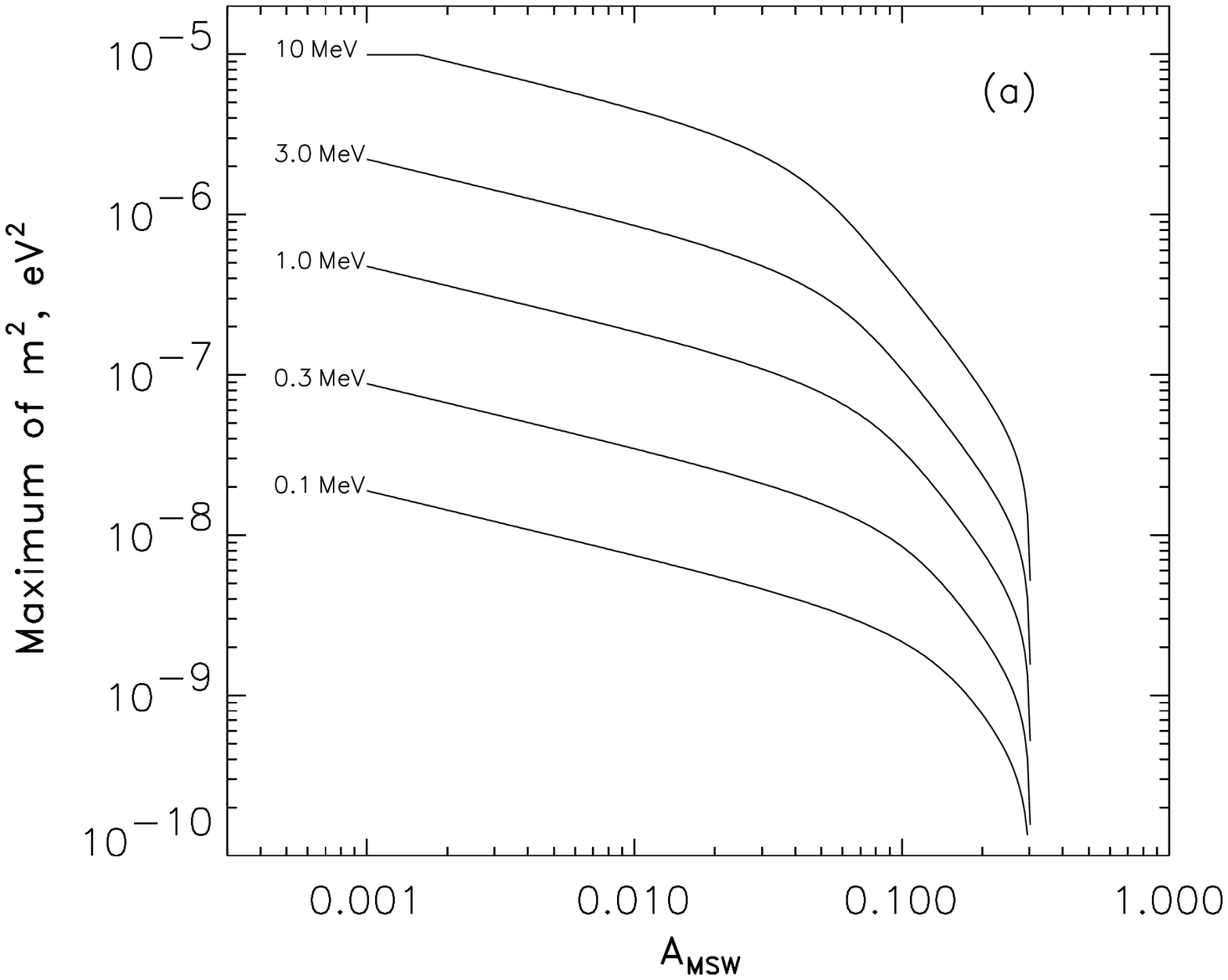}
\includegraphics{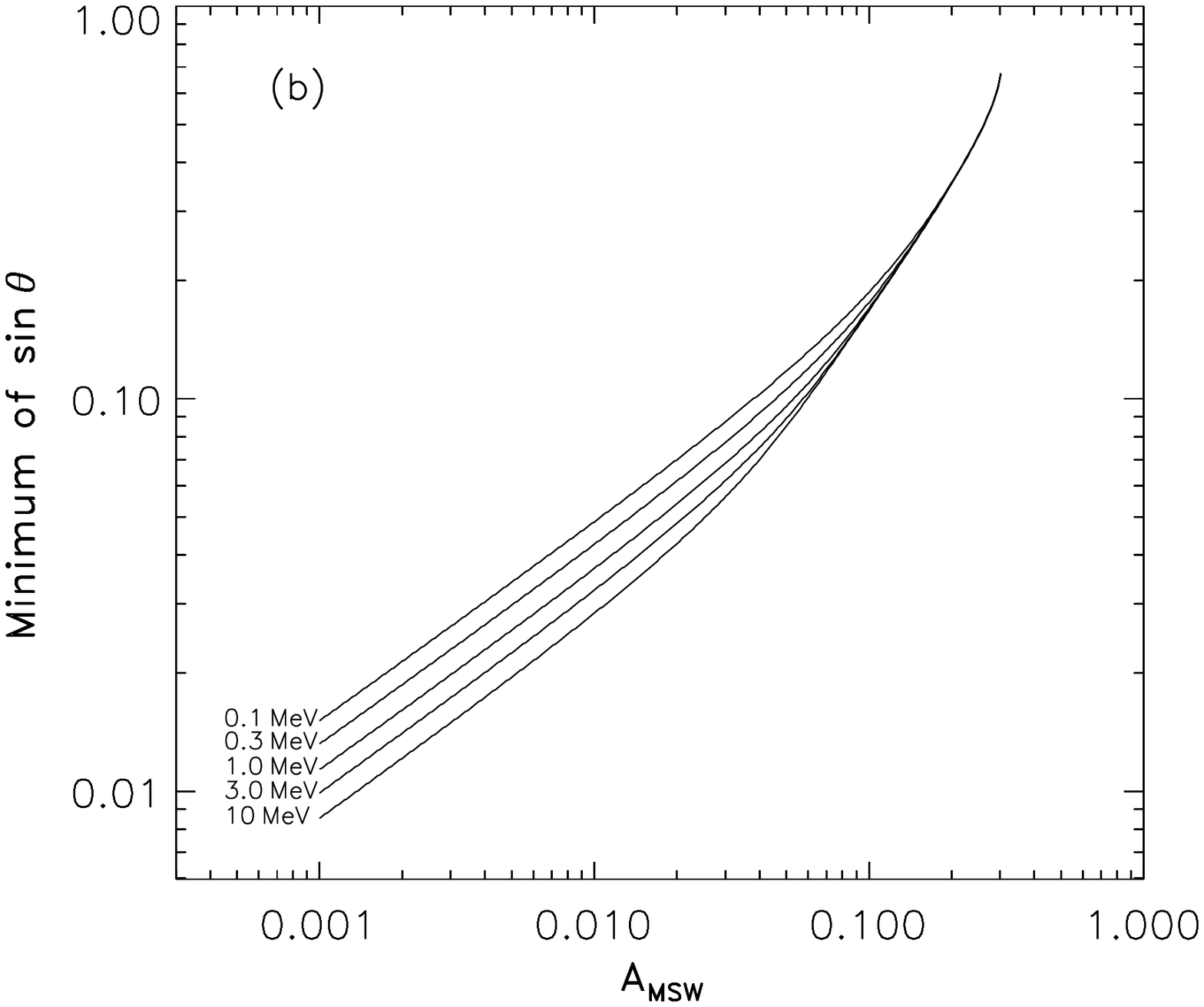}
\caption{
(a):~The maximum possible value of $m^2$ as a function of the desired level of relative neutrino flux 
variations $A_{\rm MSW}$ is represented for different neutrino energies.
(b):~The minimum possible value of $\sin\theta$ as a function of $A_{\rm MSW}$ for 
different neutrino energies.}
\label{FIGURE_CONTINUOUS_MAX_MIN}
\end{figure}

Let us assume $m^2\define m_2^2\gg m_1^2$ and use Eqs.~(\ref{JUMP_PROBABILITY}) and~(\ref{A_FACTOR})
in the same way as we did in the case of neutrino line detection. The amplitude of the neutrino flux time variations 
normalized to the standard neutrino flux, $A_{\rm MSW}$, is given by Eq.~(\ref{A_MSW_CONTINUOUS}). 
For the optimal choice of the parameter $\xi$ (i.e.~$\sigma$) it depends only on three parameters: 
${\cal E}_{\rm c}$, $m^2$ and~$\sin\theta$.
Figures~\ref{FIGURE_CONTINUOUS_ENERGY}(a),~\ref{FIGURE_CONTINUOUS_ENERGY}(b) and~\ref{FIGURE_CONTINUOUS_ENERGY}(c) 
represent contours for the relative flux variations, $A_{\rm MSW}={\rm const}$,
in the $m^2$ and~$\sin\theta$ parameter space for three values of energy, $0.1$, $1$ and $10$ Mev (solid lines). 
Figure~\ref{FIGURE_CONTINUOUS_ENERGY}(d) shows constant energy curves, ${\cal E}={\rm const}$, for the fixed
variation level $A_{\rm MSW}=10\%$ (solid lines). To understand the effect of nuclear acceleration, we plot the 
corresponding contours for the limiting case $a_{\rm H}\rightarrow 0$, as dotted lines if the effect is noticeable.
We see that nuclear acceleration makes a real difference, especially for small neutrino energies and for small 
levels of flux variation.
Figure~\ref{FIGURE_CONTINUOUS_MAX_MIN}(a) and~\ref{FIGURE_CONTINUOUS_MAX_MIN}(b) 
show the maximum possible value of $m^2$ and the minimum possible value of $\sin\theta$ respectively versus 
the desired level of the relative neutrino flux variations $A_{\rm MSW}$ for different neutrino energies.
All these figures are based on the exact formula~(\ref{A_MSW_CONTINUOUS}) and are for the optimal choice for $\sigma$.
We see again that unless neutrino masses are quite low and the mixing angle is large, we do not detect time variations 
of the neutrino flux.


\section{Conclusions}\label{CONCLUSIONS}

There are three culprits that can destroy the coherence between different neutrino paths and wipe out
the time variations of the detected neutrino flux. These are: (1)~the averaging over the region of neutrino
emission, (2)~the averaging over the effective detected energy band, and (3)~the collisional decoherence.
We concentrate on the third one in this paper. However, to properly treat the nuclear collisions, we must
also include the two other decoherence effects in our theory. 

Introducing nuclear accelerations into the theory requires a careful treatment of the process of neutrino 
emission (based on the second quantization method) and correct calculations of different neutrino emission 
times and energies. In solving this problem we use the convenient concept of the neutrino paths, with 
the two neutrino eigenfunctions as mixtures of them. Different neutrino paths have different quantum 
phases and interfere at the Earth. They also have slightly {\it different emission times} and slightly 
{\it different energies} which are determined by the method of stationary phase, which leads to the causality 
and the Doppler shift equations. Thus, there is a correlation between the flux of neutrinos at two
nearby energies. This result contradicts those of Stodolsky~\cite{S_98}, who states that for any steady source
of neutrinos there should be no such correlations. We believe that the difference arises because he neglects 
the time dependent two point correlation function for the neutrino emission. These correlations are due to
coherent fluctuations of the nuclear densities. As is well known, such fluctuations lead to correlations at 
different energies. They are produced by the nuclear accelerations we consider.

We calculate the neutrino flux time variations and the effect of nuclear collisions (best summarized in 
Fig.~\ref{FIGURE_CRITERIA}) separately for the case of a neutrino line and for the case of the continuous 
neutrino spectrum.\\
{\bf 1.}~~Solar neutrino line. We obtain the known results for this case. The averaging of the neutrino flux 
over the line Doppler broadening energy profile washes out any flux variations unless the neutrino masses are 
very small $m_2^2-m_1^2\simlt 5\times 10^{-9}\,{\rm eV}^2$ and the mixing angle is quite large 
$\sin\theta\simgt 0.3$.\\
{\bf 2.}~~Continuous solar neutrino spectrum. In order to observe flux time variations in this case, either 
neutrino masses should be very small (depending on the energy, see Fig.~\ref{FIGURE_CONTINUOUS_MAX_MIN}a), or 
the energy resolution of our detection device should be incredibly high, $\sigma\approx \sigma_t$ with 
$\sigma_t$ given by Eq.~(\ref{SIGMA_T}). See Fig.~\ref{FIGURE_CONTINUOUS_ENERGY} for detailed contours of the 
relative flux variations.

However, for both cases and for neutrino masses larger than a certain limit 
($m_2^2-m_1^2\simgt 10^{-7}\,{\rm eV}^2\,{\cal E}^{3/2}_{\rm MeV}$) nuclear collisions always destroy neutrino 
flux variations; and for masses less than this nuclear collisions are unimportant.

Recently, Bahcall et al gave three estimates for MSW solutions that best agree with experiments~\cite{BKS_98}.
Even for their low mass solution, $m_2^2-m_1^2=7.9\times 10^{-8}\,{\rm eV}^2$ and $\sin\theta=0.63$, no time 
variations of the neutrino flux from the $0.862\,{\rm MeV}$ $^7$Be line would be detected. As for the ``vacuum 
oscillation'' solutions, $m_2^2-m_1^2\sim 10^{-10}\,{\rm eV}^2$, the collisional decoherence is negligible for 
them.


\acknowledgments

We are happy to acknowledge many useful discussions of this problem with John Bahcall, Abe Loeb, Peter 
Goldreich and Marshall Rosenbluth. We also acknowledge encouragement and help from Joe Weingartner. 
This work was supported by the NASA astrophysical program under grant NAGW$2419$.



\newpage
\appendix

\section{Neutrino paths and eigenfunctions for the case when the resonance is inside the solar core.}

Let us consider that the spherical resonance region happens to be inside the region of neutrino emission.
In the simplest case a neutrino emitted outside the resonance radius does not cross the resonance at all. 
In this trivial case, we do not need to use indices $\lambda$ and $\Lambda$. There are two neutrino paths and 
two neutrino eigenfunctions which coincide with $+$ and $-$ neutrino states, and they are marked by the single 
index $\pm$.

In the more complicated case a neutrino, which has been emitted at the back side of the sun, crosses the 
resonance region twice. Because of geometrical symmetry, probabilities at the first and second resonance 
crossings for the neutrino to jump from one state to another are the same and equal to $\delta^2$. Indices 
$\lambda$ and $\Lambda$ are now three-dimensional. The four possible values for $\lambda$ are $(1,1,1)$, 
$(1,1,-1)$, $(1,-1,-1)$ and $(1,-1,1)$. The index $\Lambda$ is again the product of indices $\pm$ and $\lambda$. 
So, there are eight neutrino paths and correspondingly eight values of the path index $\Lambda$: $(+,+,+)$, 
$(+,+,-)$, $(+,-,-)$, $(+,-,+)$, $(-,-,-)$, $(-,-,+)$, $(-,+,+)$ and $(-,+,-)$. For example, the $(+,-,-)$ 
path means that a neutrino originally in the $+$ state is converted into the $-$ state at the first resonance 
crossing and stays in the $-$ state after passing the second resonance crossing. The two neutrino eigenfunctions 
coincide with $+$ and $-$ neutrino states at the point of emission and they (the eigenfunctions) are given by 
the formula
\beq
|\psi_\pm(\mbox{position of emission})\!>\:&\longrightarrow&
\:\sum\limits_\lambda {\tilde b}_\lambda |\psi_{\pm \lambda_2}(\mbox{between the two resonances})\!>\nonumber\\
\:&\longrightarrow&\:\sum\limits_\lambda b_\lambda |\psi_{\pm \lambda_3}(\mbox{vacuum})\!>,
\eeq
where the sum should be taken over all four possible values of the index $\lambda$. Here the subscripts 
$\pm\lambda_2$ and $\pm\lambda_3$ are the product of the $\pm$ index with the second and the third components of 
the vector index $\lambda$ respectively. The absolute values of the amplitudes ${\tilde b}_\lambda$ 
and $b_\lambda$ are
\beq
\begin{array}{lcl}
|{\tilde b}_{(1,1,1)}|   &=& {\Bigl(1-\delta^2\Bigr)}^{1/2},\\
|{\tilde b}_{(1,1,-1)}|  &=& 0,\phantom{{\Bigl(1-\delta^2\Bigr)}^{1/2}\delta}\\
|{\tilde b}_{(1,-1,-1)}| &=& {\delta},\\
|{\tilde b}_{(1,-1,1)}|  &=& 0,
\end{array}
\eeq
and
\beq
\begin{array}{lcl}
|b_{(1,1,1)}|   &=& {\Bigl(1-\delta^2\Bigr)},\\
|b_{(1,1,-1)}|  &=& {\Bigl(1-\delta^2\Bigr)}^{1/2}\delta,\phantom{0}\\
|b_{(1,-1,-1)}| &=& \delta{\Bigl(1-\delta^2\Bigr)}^{1/2},\\
|b_{(1,-1,1)}|  &=& \delta^2.
\end{array}
\eeq
Of course, the choice of ${\tilde b}_\lambda$ for the unused indices is arbitrary.


\section{Averaging over the position of neutrino emission}

Let us consider the case of a single resonance crossing. In this appendix we show that the cross term between 
two neutrino paths $\Lambda_1$ and $\Lambda_2$ is averaged out over the region of neutrino emission if 
these paths are different in the region of the solar core. Let for example $\Lambda_1=(-,-)$ and $\Lambda_2=(+,+)$. 
Since the phase $\phi_a a_x$ is much smaller than the phase $\phi_0$ in Eq.~(\ref{DELTA_PHI}),
we can drop it (but only for the averaging over the neutrino emission region. It still can be much larger than 
one, resulting in the collisional decoherence if the solar core paths are the same.) Then, the cross term between 
$\Lambda_2$ and $\Lambda_1$ paths is $|B_{\Lambda_1}B_{\Lambda_2}|\cos(\phi_0)$ [see Eq.~(\ref{MSW_FACTOR})]. 
Henceforward we use dimensionless variables and functions (we use the same notations for them as for dimensional 
variables):
\beq
\begin{array}{lcl}
{\bf r}={\bf r}/R_\odot & - & \mbox{radius-vector normalized to the radius of the sun,}\\
x=x/R_\odot & - & \mbox{dimensionless x-coordinate,}\\
y=r_\bot/R_\odot & - & \mbox{dimensionless perpendicular radius $r_\bot=\sqrt{r^2-x^2}$,}\\
n_e=n_e/n_e(0) & - & \mbox{electron density normalized to that for the center of the sun,}\\
\Delta=\Delta/g(0) & - & \mbox{$\Delta=(m_2^2-m_1^2)/2{\cal E}$ normalized to $g(0)=\sqrt{2}G_{\rm F}n_e(0)$,}\\
P({\bf r}) & - & \mbox{normalized neutrino production rate inside the sun.}
\end{array}
\nonumber
\eeq
It is obvious that $P({\bf r})$ is zero outside the region of the solar core and 
\beq
\int\!\!\!\int\!\!\!\int P({\bf r})\,{\bf d^3r}=4\pi\,\int\limits_0^\infty P(r)\,r^2\,dr=1.
\eeq

Now using Eqs.~(\ref{PHI_0}),~(\ref{K_PRIME}) and recalling the definition $g\define\sqrt{2}G_{\rm F}n_e$,
we get for the cross term averaged over the position of neutrino emission
\beq
{\left\langle\,|B_{\Lambda_1}B_{\Lambda_2}|\,\cos(\phi_0)\,\right\rangle}_{\bf r}=
\int\!\!\!\!\!\!\!\!\!\int\limits_{|{\bf r}|<r_{\rm core}}\!\!\!\!\!\!\!\!\!\int 
P(r)\,f({\bf r})\:
\cos{\left[A\int\limits_x^{X_{\rm E}}\,\kappa({\tilde r})\,d{\tilde x}\right]}\:{\bf d^3r}.
\eeq
Here $r_{\rm core}$ is the dimensionless radius of the solar core and we introduce
\beq
&&A\define\frac{g(0)R_\odot}{\hbar c}\approx 27000\gg1,
\label{APPENDIX_A_01}
\\
&&f({\bf r})\define |B_{\Lambda_1}B_{\Lambda_2}|,
\label{APPENDIX_A_02}
\\
&&\kappa(r)\define{\left[n^2-2n\Delta\cos{2\theta}+\Delta^2\right]}^{1/2}.
\label{APPENDIX_A_03}
\eeq
Using cylindrical coordinates $x$ and $y$, we write
\beq
{\left\langle\, |B_{\Lambda_1}B_{\Lambda_2}|\,\cos(\phi_0)\,\right\rangle}_{\bf r}=
{\rm Re}\;\int\limits_0^{r_{\rm core}} 2\pi y\,dy\:
\int\limits_{-Y}^Y P(r)\,f(x,y)\:\exp{\left[i A\int\limits_x^{X_{\rm E}}\,\kappa({\tilde r})\,d{\tilde x}\right]}\:dx,
\label{APPENDIX_A_1}
\eeq
where $Y\define\sqrt{r_{\rm core}^2-y^2}$. Using a new variable $t$ instead of $x$
\beq
t\define -\int\limits_x^{X_{\rm E}}\,\kappa({\tilde r})\,d{\tilde x},
\eeq
we rewrite Eq.~(\ref{APPENDIX_A_1}) as
\beq
{\left\langle |B_{\Lambda_1}B_{\Lambda_2}|\,\cos(\phi_0)\right\rangle}_{\bf r}=
{\rm Re}\;\int\limits_0^{r_{\rm core}} 2\pi y\,dy\:
\int\limits_{T_-}^{T_+} \frac{P(r)\,f(x,y)}{\kappa(r)}\:\exp{\left[-i A t\right]}\:dt,
\label{APPENDIX_A_2}
\eeq
where
\beq
T_\pm=-\int\limits_{\pm Y}^{X_{\rm E}}\,\kappa({\tilde r})\,d{\tilde x}.
\eeq
Integrating Eq.~(\ref{APPENDIX_A_2}) by parts and using equation~$P(r=r_{\rm core})=0$ 
(no neutrino emission outside the region of the solar core), we get
\beq
{\left\langle\, |B_{\Lambda_1}B_{\Lambda_2}|\cos(\phi_0)\,\right\rangle}_{\bf r}=
{\rm Re}\,\int\limits_0^{r_{\rm core}} 2\pi y\,dy\,
\int\limits_{T_-}^{T_+} \frac{1}{iA}\:\frac{1}{\kappa(r)}\:\frac{x}{r}\:
\frac{d}{dr}\!\left[\frac{P(r)\,f(x,y)}{\kappa(r)}\right]\,\exp{\left[-i A t\right]}\:dt.
\label{APPENDIX_A_3}
\eeq
where we also use $d/{dt}=(1/\kappa)(x/r)(d/{dr})$. To estimate the absolute value of the averaged cross term 
we drop the exponential factor in Eq.~(\ref{APPENDIX_A_3}) and we return to the old variable $x$ instead of $t$.
We have
\beq
\left|{\left\langle B_{\Lambda_1}B_{\Lambda_2}\,\cos(\phi_0)\right\rangle}_{\bf r}\right|\le
\frac{1}{A}\:\int\limits_0^{r_{\rm core}} 2\pi y\,dy\:
\int\limits_{-Y}^Y \frac{|x|}{r}\:\left|\frac{d}{dr}\frac{P(r)\,f(x,y)}{\kappa(r)}\right|\:dx.
\label{APPENDIX_A_4}
\eeq
We see that now we have a small factor $1/A\approx 4\times10^{-5}$ in the formula for the averaged cross term.
We can integrate Eq.~(\ref{APPENDIX_A_3}) by parts again and get the factor $1/A^2$, and so on.
However, instead we content ourselves with the numerical estimate from Eq.~(\ref{APPENDIX_A_4}).

Since we consider the case of a single resonance crossing, the resonance radius is outside the neutrino emission
region, $n(r_{\rm core})>\Delta$ and $\kappa(r)\approx n(r)$ [see definition~(\ref{APPENDIX_A_03})].
Let us further assume that $f(x,y)=|B_{\Lambda_1}B_{\Lambda_2}|$ is constant (as in the case when the electron 
density inside the neutrino emission region is much larger than that at the resonance radius), and replace 
$f(x,y)$ by its largest possible value $f(x,y)=1/4$, according to Eqs.~(\ref{4_B}) and for our choice 
$\Lambda_2=(+,+)$ and $\Lambda_1=(-,-)$. In this case, using the spherical coordinate system $x=r\cos\alpha$ 
and $y=r\sin\alpha$, we rewrite Eq.~(\ref{APPENDIX_A_4}) as
\beq
\left|{\left\langle B_{\Lambda_1}B_{\Lambda_2}\,\cos(\phi_0)\right\rangle}_{\bf r}\right| & \le &
\frac{\pi}{A}\:\int\limits_0^{r_{\rm core}} r^2\:\left|\frac{d}{dr}\frac{P(r)}{\kappa(r)}\right|\:dr\:
\int\limits_{0}^{\pi/2} \sin\alpha\:\cos\alpha\:\,d\alpha \nonumber\\
& = & -\frac{\pi}{2A}\:\int\limits_0^{r_{\rm core}} r^2\:\frac{d}{dr}\left[\frac{P(r)}{n(r)}\right]\:dr=
\frac{\pi}{A}\:\int\limits_0^{r_{\rm core}} r\:\frac{P(r)}{n(r)}\:dr\approx 2\times 10^{-4}.
\label{APPENDIX_A_5}
\eeq
For the final numerical estimate we use the production rate and the electron density for the standard solar 
model~\cite{B_89}. We see that an averaged cross term is indeed very small if the initial neutrino states
are different.


\section{Derivation of equation~(\ref{ACCELERATION_INTEGRAL})}

Substituting the Holtsmark function~(\ref{H}) into Eq.~(\ref{ACCELERATION_INTEGRAL}) and introducing
a new parameter $A\define\phi_a a_{\rm H}$, we get
\beq
{\left\langle\frac{\sin(\phi_a a)}{\phi_a a}\right\rangle}_a=
\frac{2}{\pi}\int\limits_0^\infty\!\int\limits_0^\infty 
\frac{\sin(A\beta)}{A\beta}\:\exp{\left[-{(x/\beta)}^{3/2}\right]}\:\frac{x}{\beta}\:\sin x\:dx\,d\beta.
\eeq
Now we change the variable $\beta$ to a new variable $t\define A\beta$ to obtain
\beq
{\left\langle\frac{\sin{(\phi_a a)}}{\phi_a a}\right\rangle}_a 
& = & \frac{2}{\pi}\int\limits_0^\infty \frac{\sin t}{t^2}\,dt
\int\limits_0^\infty \exp\left[-A^{3/2}{(x/t)}^{3/2}\right]\:x\,\sin x\,dx \nonumber\\
& = & \frac{2}{\pi}\int\limits_0^\infty \frac{\sin t}{t^2}\,dt
\int\limits_0^\infty \exp{\left[-A^{3/2}{(x/t)}^{3/2}\right]}\:\frac{d}{dx}(\sin x-x\cos x)\:dx.
\label{APPENDIX_B_1}
\eeq
We integrate the second integral in Eq.~(\ref{APPENDIX_B_1}) by parts to get
\beq
{\left\langle\frac{\sin{(\phi_a a)}}{\phi_a a}\right\rangle}_a=
\frac{3}{\pi}\:A^{3/2}\int\limits_0^\infty\!\int\limits_0^\infty
x^{1/2}\,t^{-7/2}\:\sin t\:\exp{\left[-A^{3/2}{(x/t)}^{3/2}\right]}\:(\sin x-x\cos x)\:dx\,dt.
\eeq
Introducing a new variable $y\define x/t$ instead of the variable $x$, we obtain
\beq
{\left\langle\frac{\sin{(\phi_a a)}}{\phi_a a}\right\rangle}_a&=&\frac{3}{\pi}A^{3/2}\int\limits_0^\infty
y^{1/2}\,\exp{\left[-A^{3/2}y^{3/2}\right]}\:dy \nonumber\\
&&\times\left[\,\int\limits_0^\infty\frac{\sin{(t)}\sin{(yt)}}{t^2}\,dt-
y\int\limits_0^\infty\frac{\sin{(t)}\cos{(yt)}}{t}\,dt \,\right].
\eeq
Now using formulas~\cite{GR_94}
\beq
\int\limits_0^\infty\frac{\sin(t)\,\sin{(yt)}}{t^2}\,dt = \left\{
\begin{array}{ll}
y\cdot\pi/2, & y<1\\
\pi/2, & y\ge 1
\end{array}
\right. 
\quad\mbox{and}\quad
\int\limits_0^\infty\frac{\sin(t)\,\cos{(yt)}}{t}\,dt = \left\{
\begin{array}{ll}
\pi/2, & y<1\\
\pi/4, & y=1\\
0, & y>1
\end{array}
\right. ,
\eeq
we get
\beq
{\left\langle\frac{\sin{(\phi_a a)}}{\phi_a a}\right\rangle}_a=\frac{3}{2}\:A^{3/2}\int\limits_1^\infty
y^{1/2}\:\exp{\left[-A^{3/2}y^{3/2}\right]}\:dy.
\eeq
Changing again the variable of integration to $x\define y^{3/2}$ and substituting $A=\phi_a a_{\rm H}$, we obtain 
the final result
\beq
{\left\langle\frac{\sin{(\phi_a a)}}{\phi_a a}\right\rangle}_a=
A^{3/2}\int\limits_1^\infty\exp{\left[-A^{3/2}x\right]}\:dx=
\exp{\left[-A^{3/2}\right]}=\exp{\left[-{(\phi_a a_{\rm H})}^{3/2}\right]}.
\eeq


\section{Derivation of equation~(\ref{CROSS_TERM})}

Let us introduce a parameter $A\define\sigma_a/\sigma$ and denote the integral in Eq.~(\ref{CROSS_TERM_1})
as $I(A)$,
\beq
I(A) & \define & \int\limits_{0}^1\!\int\limits_0^\infty
\exp{\left[-\frac{A^2}{4}\,\beta^2\mu^2\right]}\:H(\beta)\,d\beta\,d\mu \nonumber\\
& = & \frac{2}{\pi}\int\limits_0^1\!\int\limits_0^\infty\!\int\limits_0^\infty
\exp{\left[-\frac{A^2}{4}\,\beta^2\mu^2\right]}\:
\exp{\left[-{(x/\beta)}^{3/2}\right]}\:\frac{x}{\beta}\:\sin x\:dx\,d\beta\,d\mu.
\eeq
Changing the variable $x$ for a new variable $y\define x/\beta$, we get
\beq
I(A)=\frac{2}{\pi}\int\limits_0^1 d\mu  \int\limits_0^\infty y\exp{\left[-y^{3/2}\right]}\,dy\:
\int\limits_0^\infty \beta\:\exp{\left[-{\left(\frac{A\mu}{2}\right)}^2\,\beta^2\right]}\:\sin{(\beta y)}\:d\beta.
\eeq
Using the standard formula for the integral over $\beta$~\cite{GR_94}, we obtain
\beq
I(A)=\frac{4}{\pi^{1/2}\,A^3}\,\int\limits_0^\infty y^2\:\exp{\left[-y^{3/2}\right]}\:dy\:
\int\limits_0^1 \exp{\left[-\frac{y^2}{A^2}\,\frac{1}{\mu^2}\right]}\:\frac{d\mu}{\mu^3}.
\eeq
Now introducing a new variable $t\define 1/\mu^2$ instead of $\mu$, we get
\beq
I(A)=\frac{2}{\pi^{1/2}\,A^3}\int\limits_0^\infty y^2\exp{\left[-y^{3/2}\right]}dy
\int\limits_1^\infty \exp{\left[-\frac{y^2}{A^2}\,t\right]}dt=
\frac{2}{\pi^{1/2}\,A}\int\limits_0^\infty \exp{\left[-y^{3/2}-\frac{y^2}{A^2}\right]}dy.
\eeq
Changing the variable of integration to $x\define y^{1/2}$ and substituting $A=\sigma_a/\sigma$, we obtain 
the final result
\beq
I(\sigma_a/\sigma)=\frac{4}{\pi^{1/2}}\,\frac{\sigma}{\sigma_a}\,\int\limits_0^\infty x\:
\exp{\left[-x^3-x^4\,\frac{\sigma^2}{\sigma_a^2}\right]}\:dx.
\eeq


\phantom{0}\newpage\phantom{0}

\phantom{0}\newpage\phantom{0}

\phantom{0}\newpage\phantom{0}

\phantom{0}\newpage\phantom{0}

\phantom{0}\newpage\phantom{0}

\phantom{0}\newpage\phantom{0}

\end{document}